\documentclass[12pt,preprint]{aastex}

\newcommand {\sw}{{\it Swift}}
\newcommand {\xmm}{{\it XMM--Newton}}
\newcommand {\nh}{{N$_{\rm H}$}}
\newcommand {\gsim}{ \lower .75ex \hbox{$\sim$} \llap{\raise .27ex \hbox{$>$}} } 
\newcommand {\lsim}{ \lower .75ex\hbox{$\sim$} \llap{\raise .27ex \hbox{$<$}} } 

\usepackage{graphicx}
\usepackage[hang]{subfigure}
\graphicspath{{./}}

\usepackage[normalem]{ulem}  

\shorttitle{Three years {\em Swift}--BAT survey of AGN}
\shortauthors{Burlon et al.}

\begin{document}
\title{Three years {\em Swift}--BAT Survey of AGN: reconciling theory and observations?\altaffilmark{*}}


\author{D. Burlon\altaffilmark{1},\email{burlon$@$mpe.mpg.de}
 M. Ajello\altaffilmark{2},\email{majello$@$slac.stanford.edu}
J. Greiner\altaffilmark{1},
A. Comastri\altaffilmark{3},
A. Merloni\altaffilmark{1,4}, and
N. Gehrels\altaffilmark{5}
}


\altaffiltext{1}{Max-Planck-Institut f\"ur Extraterrestrische Physik, 
Giessenbachstra\ss{}e 1, 85740 Garching, Germany }

\altaffiltext{2}{SLAC National Laboratory and Kavli Institute for 
Particle Astrophysics and Cosmology, 2575 Sand Hill Road, Menlo Park, CA 94025, USA}

\altaffiltext{3}{INAF Osservatorio Astronomico di Bologna, via Ranzani 1, 40127 Bologna, Italy}

\altaffiltext{4}{Excellence Cluster Universe, TUM, Boltzmannstra\ss{}e 2, 85748, Garching, Germany}

\altaffiltext{5}{Astrophysics Science Division, Mail Code 661, NASA Goddard Space Flight Center, Greenbelt, MD 20771, USA}

\altaffiltext{*}{Based on observations obtained with XMM-Newton, an ESA science mission with instruments 
and contributions directly funded by ESA Member States and NASA.}

\begin{abstract} 
It is well accepted that unabsorbed as well as absorbed AGN are needed to explain the nature and the shape of the Cosmic X-ray background, even if the fraction of  highly absorbed objects (dubbed Compton-thick sources) substantially still escapes detection. We derive and analyze the absorption distribution  using a complete  sample of AGN detected by \sw--BAT in the first three years of the survey. The fraction of Compton-thick AGN represents only 4.6\% of the total AGN population detected by \sw--BAT. However, we show that once corrected for the bias against the detection of very absorbed sources the real intrinsic fraction of Compton-thick AGN is 20$^{+9}_{-6}$\,\%. 
We proved for the first time (also in the BAT band)
that the anti-correlation of the fraction of absorbed
AGN and luminosity it tightly connected to the
different behavior of the luminosity functions (XLFs) of absorbed and 
unabsorbed AGN.
This points towards a difference between the two subsamples of objects with absorbed AGN being, on average, intrinsically less luminous than unobscured ones. Moreover the XLFs show that the fraction of obscured AGN might  also decrease at very low luminosity. This can be successfully interpreted in the framework of a disk cloud outflow scenario as the disappearance of the obscuring region below a critical luminosity. Our results are discussed in the framework of population synthesis models and the origin of the Cosmic X-ray Background.
\end{abstract}




\section{Introduction}
It is well known that absorbed active galactic nuclei (AGN) are needed to
explain the shape of the Cosmic X-ray background (CXB) spectrum
\citep[e.g.][]{gilli07,treister09}. A large
fraction of them is indeed detected in the shallow and deep $<10$\,keV
X-ray surveys \cite[see e.g.][]{brandt05}. 
Nonetheless, a large fraction of Compton-thick (N$_{H}\geq1.5\times10^{24}$\,H-atoms cm$^{-2}$, for a review see \citealp{comastri04}) 
AGN still escapes detection.
Because of  their large absorbing column density, 
these sources contribute  a $\sim$10--25\%
fraction of the CXB emission \citep[][respectively]{gilli07,treister09} at 30 keV, but 
at the same time are expected to be fairly numerous representing
up to $\sim30$\,\% of the entire AGN population \citep[][]{risaliti99,worsley05} .
The advent of sensitive all-sky surveys above 15\,keV (e.g. \sw--BAT
and INTEGRAL) opened the possibility to detect these objects. Indeed,
it is above 10--15\,keV that part of the primary continuum emission
pierces through the veil of Compton-thick material, making it easier
for these objects to be detected. Despite this fact, the early results from
the 	\sw\ and INTEGRAL surveys showed that the fraction of Compton-thick
AGN is merely a 5--10\% of the total AGN population \citep[see][
and references therein]{ajello09c}.  
Thus it might be that the fraction of  Compton-thick AGN is intrinsically smaller than previously determined. 
Nevertheless, it should be taken into account that even above $>15$\,keV, instruments are biased against the detection
of logN$_H>24$ sources. Indeed as shown in e.g. \cite{ghisellini94}, \cite{ajello09c},
50\% of the source flux (between 15-55 keV) is lost if logN$_H>24.5$. 
The fact that three of the closest AGN ever detected
(i.e. NGC 1068, Circinus galaxy and NGC 4945) are absorbed by N$_{H}\geq10^{24}$\,atoms cm$^{-2}$ 
suggests that indeed this is the likely explanation.

In this work, we present and discuss the results of the most complete -~until~now~- survey
of AGN in the local Universe (z$<$0.1) using data from the \sw--BAT telescope.
We performed a detailed spectral study of the three years sample of Seyfert-like objects by combining 
the hard X-ray information with the soft X-ray observations realized by different missions in the recent past.
To this aim we extracted \sw--BAT spectra and spectra in the 0.3--10 keV band using
archival \xmm\ and \sw--XRT data.
In a handful of cases we requested and obtained target-of-opportunity (ToO) 
observations with \sw\ for objects without previous coverage at soft X-rays.
The use of X-ray data in the 0.3--195\,keV band allows us to constrain robustly
 all the source parameters (including the absorbing column density).

This paper is organized as follows. In \S2 we  present the sample and discuss how the joint spectral  analysis  was performed.
The general properties of the 15--195\,keV continuum of AGN, 
the stacked analysis for the absorbed, unabsorbed, and Compton-thick spectra are presented in \S3.
\S4 presents the observed \nh\ distribution. 
We then evaluate, for the BAT survey, the bias against the  detection of the most 
obscured AGN and estimate -for the first time- the intrinsic absorption distribution. 
\S5 presents the anti-correlation between the fraction of absorbed AGN (relative
to the whole population) and luminosity, while the luminosity functions
of AGN are derived in \S6.
The results of these analyses are discussed in \S7 while \S8 summarizes our findings.
In this work we use a standard cosmology (H$_0$ = 70, q$_0$ = 0, and $\Omega_{\Lambda}$ = 0.73).

\section{The {\em Swift}--BAT sample and data analysis}

\subsection{The sample} 
The Burst Alert Telescope \citep[BAT;][]{barthelmy05}
onboard the {\em Swift} satellite \citep{gehrels04}, represents 
a major improvement in sensitivity for imaging of the hard X-ray sky. 
BAT is a coded mask telescope with a wide field of view 
(FOV, 120$^\circ\times$90$^{\circ}$ partially coded) aperture, sensitive in
the 15--200\,keV domain. Thanks 
to its wide FOV and its pointing strategy, BAT monitors continuously
up to 80\% of the sky every day achieving, after several years of the
survey, deep exposure in the entire sky.
Results of the BAT survey \citep{markwardt05,ajello08a,tueller10}
show that BAT reaches a sensitivity of $\sim$1\,mCrab\footnotemark{}
\footnotetext{1\,mCrab in the 15--55\, keV band corresponds to 
1.27$\times10^{-11}$\,erg cm$^{-2}$ s$^{-1}$}
in 1\,Ms of exposure. 
Being the BAT survey not a flux-limited survey, but rather a significance-limited one, it is important to 
address how the survey flux limit changes over the sky area. This is often referred to as sky coverage, 
that is the distribution of the surveyed area as a function of limiting flux. Its knowledge is very important 
when performing population studies as the ones described in the next sections. The reader is referred to \cite{ajello08a} 
for how to derive the sky coverage as a function of the minimum detectable flux F$_{min}$. This is defined as 
the sum of the area covered to fluxes f$_i$ $<$ F$_{min}$:
\begin{equation}
\Omega(<F_{min}) = \sum^{N}_{i} A_i  \ \ \\ , f_i<F_{min}
\end{equation}
where N is the number of image pixels and A$_i$ is the area associated to each of them. A visual representation 
of the sky coverage is reported in \cite{ajello09b} 
which shows clearly the good sensitivity of BAT.  
 The survey, in our analysis (15--55 keV),  reaches a limiting sensitivity of  $\sim$0.6 mCrab ($7.3\times 10^{-12}$ erg cm$^{-2}$ s$^{-1}$ ).
 We {\it ex post} checked that cutting the sample at 50\,\% of the complete sky coverage (i.e. at $1.1\times 10^{-11}$ erg cm$^{-2}$ s$^{-1}$) 
does not affect significantly the findings of this work, since the 35 objects below this limit do not populate a particular \nh\ range.

The sample used in this work is the collection of non-blazar AGN detected
by BAT during the first three years, more precisely between March 2005
 and March 2008. This sample is part of the one used in \cite{ajello09b}
which comprises all sources detected by BAT at  high ($|$b$|$$>$15$^{\circ}$)
Galactic latitude and with a signal-to-noise ratio (S/N) exceeding 5\,$\sigma$. 
All the 199 sources which are identified
as non-blazar AGN (e.g. Seyferts)  constitute the sample used in this work
and are reported (along with their properties) in the table at the end of the paper. 
Note that the main sample,
from which the sub-sample of AGN is derived, comprises 307 objects
of which only 7 are as of today without identification.
The incompleteness or the parent population is thus 2.3\,\%. 
We note that in our sample 19 objects are classified as 'Galaxies'
or 'Liners'. We believe these are normal AGN (e.g. Seyfert galaxies) for
which an accurate optical classification is not yet available in the literature.
This is based on the fact that the average redshift, luminosity and 
absorbing column density are respectively 0.03, 4.5$\times10^{43}$erg s$^{-1}$
and $10^{23}$\,cm$^{-2}$. These values are in good agreement with
the ones derived from the rest of the sample giving confidence to our hypothesis
that these objects are AGN.


\subsection{Extraction of {\it Swift}--BAT spectra}
For each source in our sample we extracted a 15--195\,keV 
spectrum  following the method described in \cite{ajello08b}. Here we recall the main steps: the details can be
found in the aforementioned paper. For a given source, we extract a 12 channel spectrum from each 
observation where the source is in the field of view. These spectra are corrected for 
residual background contamination and for vignetting; the per-pointing spectra are then (weighted) averaged to 
produce the final source spectrum. Thus, the final spectrum represents the average source emission 
over the time-span considered here (three years). 
Moreover the reader is referred to  \cite{ajello09a} for
a discussion about the accuracy of the spectra extracted with this method.

\subsection{Extraction of the soft X-ray spectra}
The goal of the present work is to obtain a reliable estimate of the intrinsic 
absorbing column density  for the BAT AGN. 102 objects (out of the 199 AGN in
our sample) have a reliable estimate of the absorbing column density present in
the literature.
The large majority (86) of these N$_{H}$ measurements  comes from earlier results of the
\sw\ survey \citep{tueller08, winter08, winter09, winter09b}  
while for the rest (16) we used single-source publications (see the table at the end of the paper for details).

For all the  objects without a \nh\ measurement we used the available follow-up observations 
performed by two observatories (i.e. \sw--XRT and \xmm\ in  83 and 12 cases, respectively). 
XRT was used preferentially, while
\xmm\ was used in a handful of cases (i.e. when the detection significance by \sw--XRT was too low to constrain  the spectral parameters and/or the \nh). Only in  2 cases we could not find any XRT or XMM follow up (i.e.
1H 2107-097) or the available soft X-ray pointing was not deep 
enough to extract a spectrum of the source (i.e. [VV2003c]~J014214.0+011615).

For the filtering and spectra extraction we used {\texttt{xrtproducts}} only on Photon Counting Level 2 
event files (grades 0--12) and the standard ftools of {\it{Headas v6.6.3}} software, and {\it{SAS v9.0.0}} 
for \sw--XRT and \xmm\ observations respectively. 
We used $Xspec$ $v12.4.0ad$ \citep{arnaud} to perform, for each AGN, joint spectral fits 
between the 15--195\,keV and the 0.3--10\,keV data.
Normally the spectra of sources detected (in the 0.3--10\,keV band) with sufficient S/N
were re-binned as to have a minimum of  20 counts/bin.
In  a handful of cases spectra were re-binned as to have 10 counts/bin, and consequently Cash in place of $\chi ^2$ statistics was adopted.

During the spectral fitting stage we took into account, regardless of the spectral model used (i) the local Galactic absorption \citep{kalberla} 
and (ii) a normalization factor to account for the different 
inter-calibration of the two instruments and for a possible variation  of the source
between the observation epochs.  Small differences in the computed value of \nh\ might be present when comparing observations taken at different times. The case of NGC~7582 is discussed in \S4.2, but in general we warn that variations of the column densities are expected \cite[see e.g.][and references therein]{risaliti02,risaliti09,bianchi09}. 

\section{General properties}
\begin{figure}
\resizebox{\hsize}{!}{\includegraphics[]{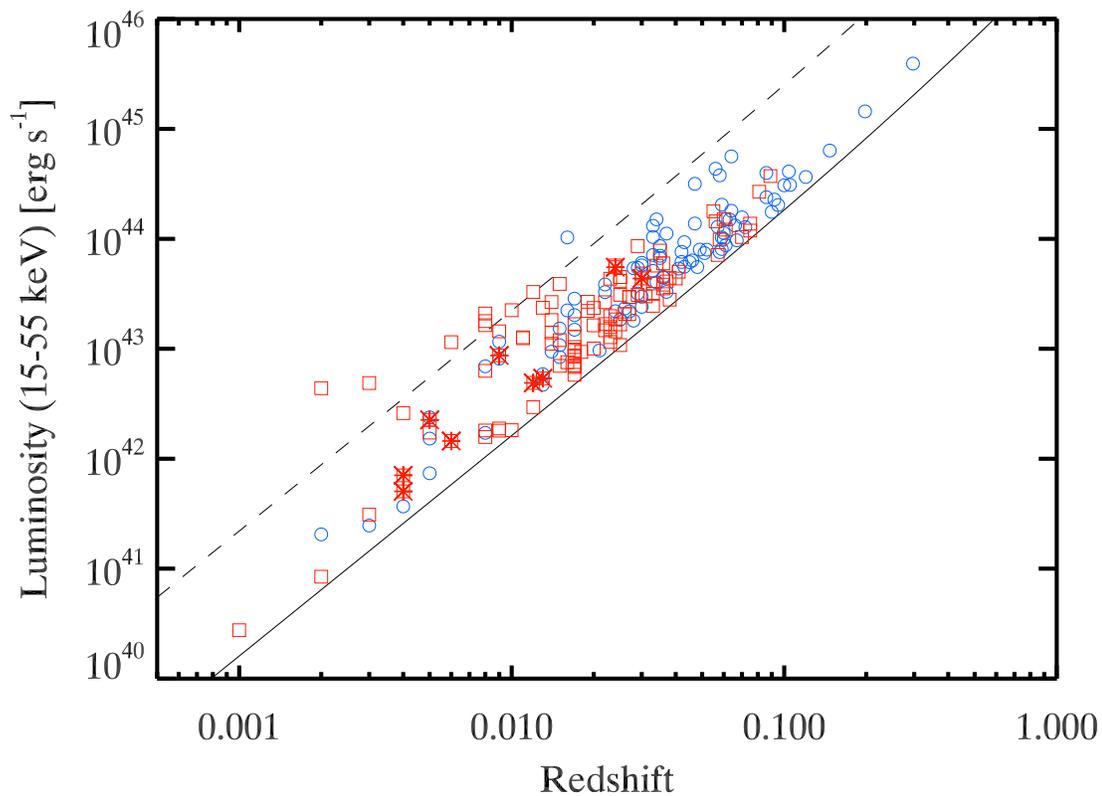} }
\caption{Luminosity-redshift plane for unabsorbed sources (blue circles) and for absorbed 
sources (red squares). The solid line represents the limiting sensitivity of the \sw--BAT survey for a source with photon index equal to 1.9, while the 
dashed line represents the sensitivity of the survey at a brighter flux, i.e. $10^{-10}$ erg cm$^{-2}$ s$^{-1}$. } 
\label{lz}
\end{figure} 

We discuss in the following the general properties of the sample, focussing on
the hard X-ray continuum emission. The joint analysis is considered in this
section only to the aim of splitting the parent population in un-absorbed and absorbed 
sources.
Fig.~\ref{lz} shows the luminosity-redshift plane for all the AGN in the BAT sample.
The  $k-$corrected $L_X$ luminosities  (not corrected for absorption, but see \S3.2 and \S4.2) were computed according to:
\begin{equation}
L_X\, =\, 4\pi d_{\rm L}^2 {F_X \over (1+z)^{2-\Gamma_X} } 
\label{lx}
\end{equation}
where $F_X$ is the X-ray flux in the 15--55 keV energy range
as listed in the table at the end of the paper 
\citep[see][for details about the flux determination]{ajello08c}, and $\Gamma_X$ is the photon 
spectral index.
Throughout this work, absorbed sources are those with an absorbing column density (N$_{\rm H}$)
larger than (or equal to) 10$^{22}$\,atoms cm$^{-2}$. It is apparent from Fig.~\ref{lz} that obscured
AGN populate more densely the low-luminosity/low-redshift part of the graph with
respect to the high-luminosity/high-redshift part.
A Kolmogorov-Smirnov (KS) test between  the redshifts of the two populations of AGN (absorbed and unabsorbed)
shows that the probability that both classes are drawn from the same parent population
is $\sim 1.4\times 10^{-3}$. 

The lines reported in Fig.~\ref{lz} represent the current limiting flux of the BAT survey
($\sim 7.3 \times 10^{-12}$ erg cm$^{-2}$ s$^{-1}$, solid line)
and a much brighter flux of $10^{-10}$ erg cm$^{-2}$ s$^{-1}$ (dashed line). 
In the shallower case it is apparent 
that sources detected are preferentially absorbed with log\nh\ larger than 22.
\begin{figure}
\resizebox{\hsize}{!}{\includegraphics[]{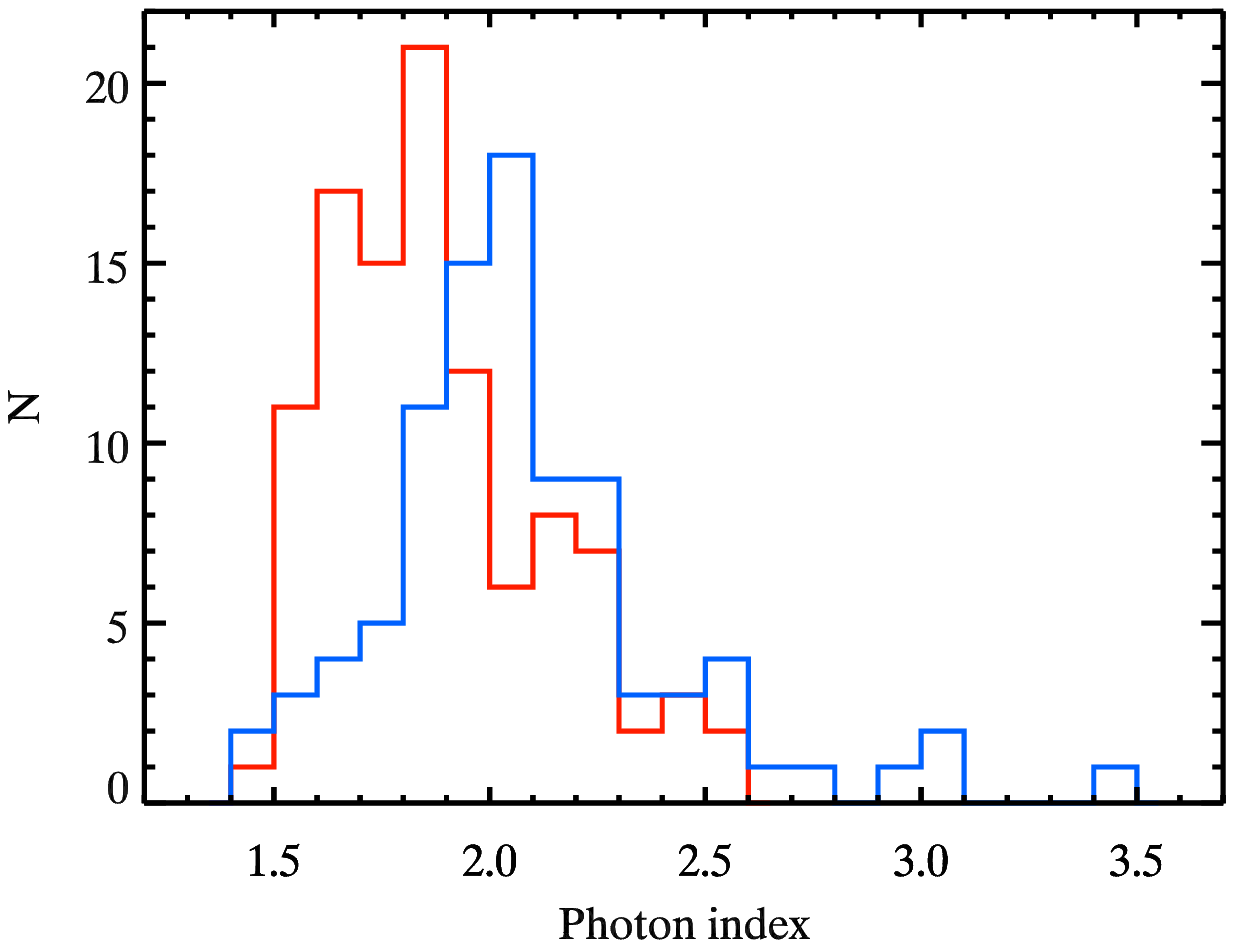} }
\caption{Photon indices distribution of absorbed AGN (logN$_H\geq$22, red line)
and unabsorbed AGN (logN$_H<$22, blue line). The photon indices used
are the ones derived in the BAT band only (i.e. 15--195\,keV band)
and reported in the table at the end of the paper. 
\label{fig:agn_index}}
\end{figure}

\subsection{Analysis of the 15-195\,keV continuum}
\begin{figure}
\resizebox{\hsize}{!}{\includegraphics[]{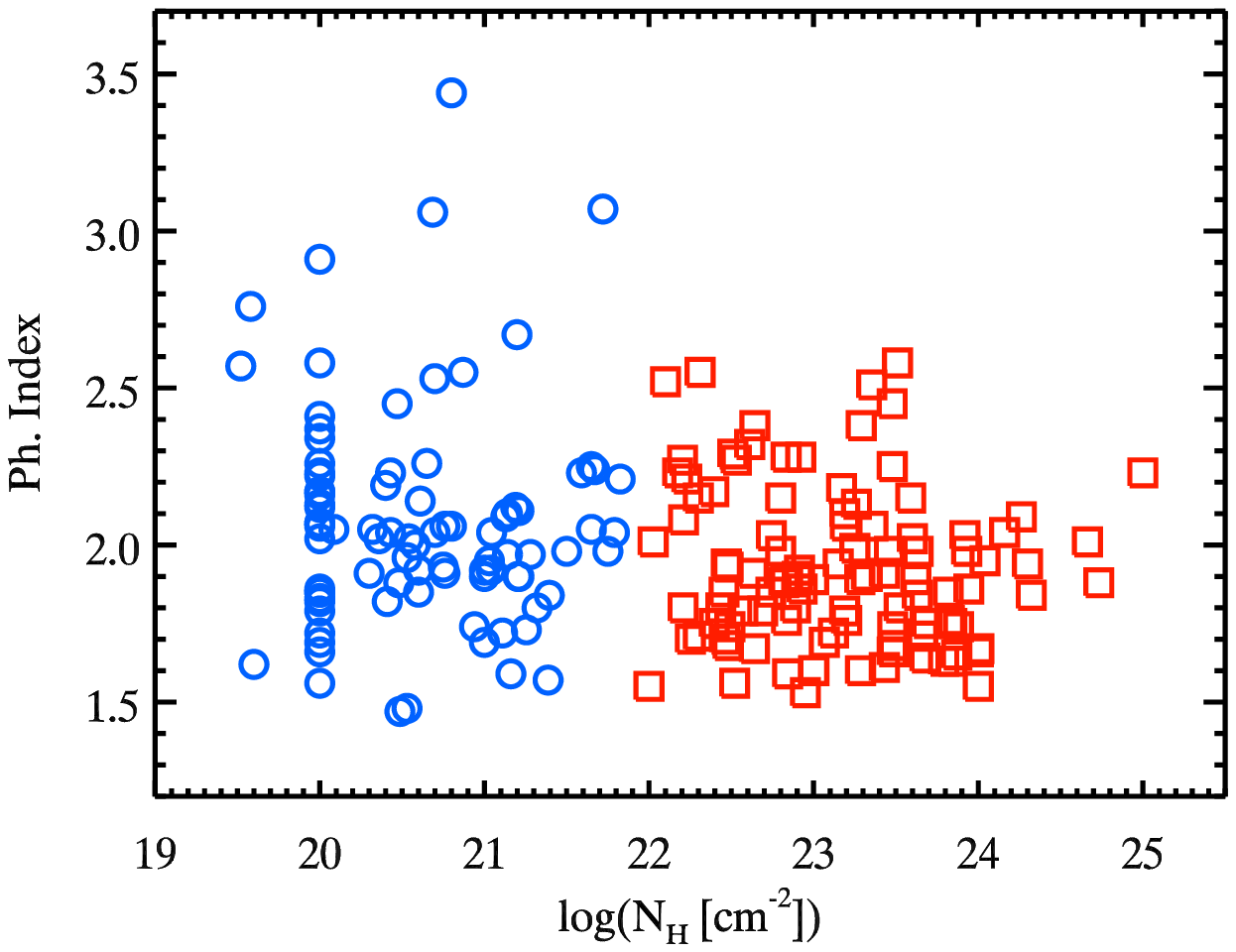} }
\caption{  Photon spectral index as obtained from fitting with a simple power-law the BAT spectrum alone, versus the \nh\
found from the combined spectra in the 0.3--195 keV energy range. Different color/symbol identify absorbed and unabsorbed sources.} 
\label{alfanh} 
\end{figure}

We performed a simple power-law fit to the BAT spectra in the 15--195\,keV band
to derive a measurement of the photon index.
These values are reported in the last table 
with the corresponding 90\% uncertainties.
The distribution of photon indices of all the BAT AGN is compatible with a Gaussian
distribution with a mean of 1.95$\pm0.02$ and a sigma of 0.27$\pm0.02$.
We analyzed separately the distribution of photon indices
of obscured and unobscured sources respectively. These
are reported in Fig.~\ref{fig:agn_index}. We find that
the two distributions appear to be different with the one
of obscured sources showing a mean of 1.92$\pm0.02$ and a sigma of 0.25$\pm0.02$
while the one of unobscured AGN displays a mean of 2.07$\pm0.03$ and a sigma
of 0.27$\pm0.03$.
This is also confirmed by the Kolmogorov-Smirnov test
which  yields a probability of $3\times10^{-3}$ that the two
distributions are drawn from the same parent population. 
 The CT sources are not the main drivers of the distribution of absorbed sources. Indeed eliminating the 9 most absorbed sources from the sample does not in turn introduce an appreciable difference in the distribution. The KS test returns a null probability of $\sim1\times10^{-3}$ in this case.
Thus, there is an indication, albeit marginal,
that absorbed sources display, on average, a harder power-law continuum
than unobscured ones. 

According to \cite{hopkins09} radiatively inefficient accretion flows (RIAFs)
make the intrinsic  X-ray spectrum of an AGN harder. This may cause
to incorrectly classify, at energies below $<$10\,keV,
 an AGN as obscured if only simple estimators
(e.g. hardness ratios) are used.  This is not however the case for 
the present work, because: 1) BAT is able to sample the intrinsic
power-law spectral index independently of the level of absorption, 2) all
sources have sufficient signal-to-noise ratio to correctly derive
the absorption level using \xmm\ or \sw--XRT data in conjunction
with BAT ones. Thus, while RIAFs might certainly affect the intrinsic
shape of the 15-195\,keV continua, we believe that the differences observed here
are ascribed to orientation effects \citep[as shown already in][]{ajello08b}.
As it can be seen in Fig.~\ref{alfanh} we show the scatter plot of photon 
 indices versus the absorbing column density and  indeed there is  a weak indication of a correlation between the two parameters (the Spearman's rank is -0.27, and null hypothesis probability $P\sim1.2\times10^{-4}$). The low, negative rank correlation coefficient and the $P$ value indicate that a chance correlation can be excluded at more than the $\gsim3\sigma$ level. However if the sources with unconstrained \nh\ are excluded from the sample  the correlation is not statistically significant any longer ($P=1.3\times10^{-3}$). A mild correlation is anyway expected, because of the contribution of the higher reflection component of type 1 AGN, in the low energy channel of the BAT. The ``softening'' effect that is introduced when fitting with a simple power-law is further discussed in the following section.

We note that three AGN (i.e.  Mrk~766, IRAS~05480+5927, Mrk~739) show a very soft BAT spectrum (e.g. photon index $>$3). 
For Mrk~766 we analyzed XRT and BAT data jointly and found out that the intrinsic power-law seems to be $\sim$2.0 and that indeed a large reflection component is required (the data show also the presence of a soft excess). The large reflection component is what makes (very likely) the BAT spectrum softer. 
The BAT spectrum of IRAS~05480+5927 is quite noisy and very soft. Nonetheless by a joint fit with XRT the photon index is constrained to be $\sim$1.8. A cutoff-powerlaw is statistically (ftest probability $>$ 3$\sigma$) better but in turn requires the cutoff energy to be at 18 keV (in the 12--30 keV range, 3$\sigma$ contours). 
As for Mrk~739, a joint fit with XRT gives a slope of 1.7. We note that again a cutoff-powerlaw is statistically preferred with a cutoff energy of $\sim50$ keV.
Our results are unaffected by the change of these three BAT-soft spectra to the flatter value reported for the joint analysis. 

\begin{table}[ht]
\centering
\scriptsize
\caption{Best-fit parameters for the stacked spectra. Errors are 90\,\% CL and parameters without an error were kept fixed during the fitting stage.
The columns report the value of the photon index ($\Gamma$), 
normalization of the reflection component $R$, the cut-off energy $E_c$
and the absorption.}

 \label{tab:stack}
\begin{tabular}{lcc cccc}

\hline
\hline
SAMPLE  & \# Obj  & $\Gamma$ & $R$ & $E_c$ & $N_{\rm H}$ & $\chi^2/dof$  \\
        &             &          &     &  (keV) & (10$^{22}$\,cm$^{-2}$) & \\
\hline

All           &  199 &  1.78$^{+0.25}_{-0.39}$ & $<4.50$ & $>80$ & -- & 5.2/8\\
All           &  199 &  1.80$^{+0.08}_{-0.08}$           & 1.00$^{+0.48}_{-0.36}$ & 300 & -- & 5.2/9\\

Absorbed$^1$      & 96   & 1.74$^{+0.07}_{-0.07}$ & 0.55$^{+0.67}_{-0.35}$  & 300 & -- & 3.3/9\\

Unabsorbed$^1$    & 92   & 1.71$^{+0.10}_{-0.06}$ & 1.23$^{+1.12}_{-1.00}$ & 300 & -- &3.6/9\\

Compton-thick &  9      & 1.80  & -- & 88$^{+35}_{-21}$ & 265$^{+171}_{-131}$ & 9.6/9\\ 

\hline

\end{tabular}
\begin{list}{}{}
\item $^{\mathrm{1}}$ For absorbed and unabsorbed AGN we have assumed that
the inclination angle between the normal to the disk/torus and the line of
sight is 60 and 30 degrees respectively. The class of absorbed AGN includes
all AGN which are absorbed but are not Compton-thick (e.g. 22$\leq$LogN$_{\rm H} \leq$24).
We did not include the 2 AGN for which \nh\ could not be calculated.
\end{list}
\end{table}

\begin{figure}
\resizebox{\hsize}{!}{\includegraphics[]{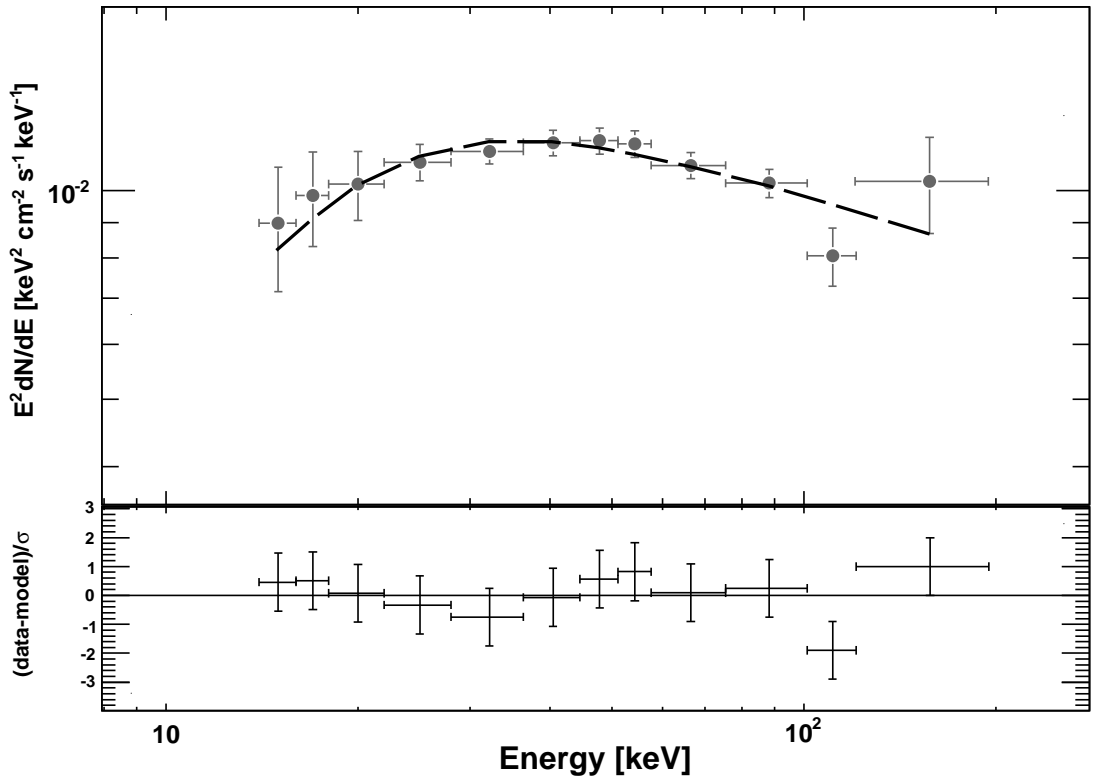} }
\caption{Stacked spectrum of the 199 Swift AGN. The dashed line
is the best-fit PEXRAV model. Note the substantial curvature of the
spectrum around $\sim$30--40\,keV.} 
\label{fig:agn_stack} 
\end{figure}

\emph{Stacked spectra analysis} -- 
In order to investigate the global spectral  properties of AGN
we performed a stacking analysis of the AGN in the BAT sample.
The stacked spectrum of several sources is produced performing the weighted average of all the spectra. The weight is chosen to be the inverse of the variance of a given bin and it is exactly the same procedure used to extract the spectra of each individual source. The same stacking technique has been already applied  with success to both the study of Seyfert galaxies and galaxy clusters detected by BAT \citep{ajello08c,ajello09a}. This stacking technique is appropriate for the stacking of background-subtracted spectra generated by coded masks telescopes. As already reported in \cite{ajello09a}, this stacking technique allows to determine  the average properties of a source population.
The stacked spectrum of the 199 AGN is not compatible with a simple
power law ($\chi^2$/dof = 22.62/10). This is due to a substantial curvature
of the spectrum around 30\,keV (see Fig.~\ref{fig:agn_stack}).
We found that an acceptable fit to the data  ($\chi^2/dof=5.2/8$)
is achieved when using a PEXRAV model. In this case we find that the 
best fit parameter for the slope is  1.78$^{+0.25}_{-0.39}$ (error
are 90\,\% CL). Given the small dynamic range of the BAT spectrum 
(15--195\,keV) is impossible to disentangle uniquely the reflection component
and the cut-off energy. Indeed, our best fit shows that, at 90\,\%
confidence, the normalization of the reflection component is consistent
with zero while the cut-off energy is bound to be $\geq80$\,keV.
The parameters of this best fit are reported in Tab.~\ref{tab:stack}.
In order to avoid this degeneracy we fixed the cut-off energy at 300\,keV (see \citealp{dadina08}, which reported an average cut-off of 300 keV for {\it BeppoSAX} sources).
The best fit parameters (reported also in Tab.~\ref{tab:stack})
for the photon index and the reflection are respectively 1.80$^{+0.08}_{-0.08}$   
and 1.00$^{+0.48}_{-0.36}$, which are in good agreement
with the findings of \cite{nandra94}. The normalization of the reflection component is 
compatible with the presence of a reflecting medium which
covers an angle of $2\pi$ at the nuclear source.

In addition we generated stacked spectra for unabsorbed (logN$_{\rm H}<22$),
absorbed (22$\leq$logN$_{\rm H}\leq24$) and Compton-thick 
(logN$_{\rm H}>24$) AGN.
 A simple power-law fit to the spectra of unabsorbed and absorbed AGN yields that the best-fit
photon indices are 2.13$\pm0.06$ and 2.00$\pm$0.06 respectively. This is found to be in agreement
with what derived from the analysis of the photon index of the two distributions: i.e. on average unabsorbed
AGN have steeper spectra than absorbed ones. The indices derived from the stacking analysis are slightly
steeper than the average ones derived from the photon index distribution because the stacked spectra show a significant curvature which makes the simple power-law fit not the  most accurate one (e.g. reduced $\chi^2\geq2.0$). We thus decided to fit the stacked spectra with a PEXRAV model.
Also in the stacked spectra of absorbed and unabsorbed AGN
the cut-off energy and the reflection component cannot be determined
uniquely. We thus fixed the cut-off energy to 300\,keV.
A fit to the stacked spectrum of absorbed sources yields
a photon index of 1.74$^{+0.07}_{-0.07}$, and
a reflection component of R=0.55$^{+0.67}_{-0.35}$.
\begin{figure}
\resizebox{\hsize}{!}{\includegraphics[]{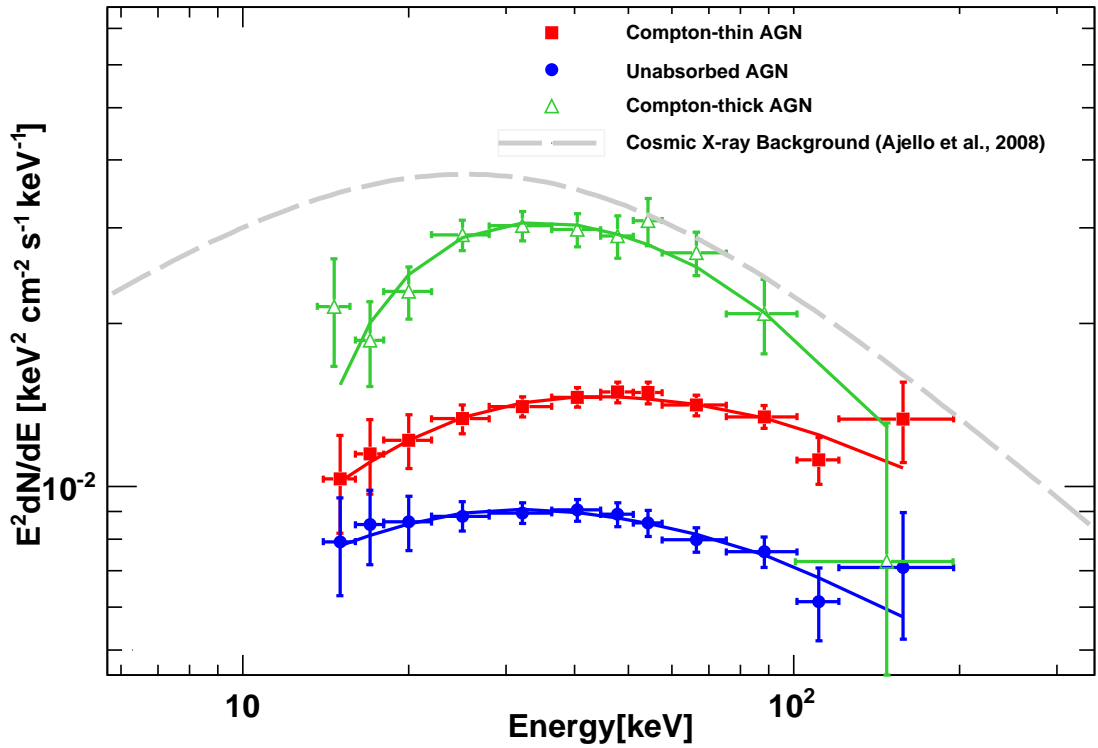} }
\caption{Stacked spectra of obscured AGN ($22\leq$logN$_{\rm H} \leq 24$),
unabsorbed AGN (logN$_H<22$) and Compton-thick AGN (logN$_{\rm H}>24$)
compared to the Cosmic X-ray background (CXB) as measured by \cite{ajello08c}. 
Note that both the CXB spectrum and the spectrum of Compton-thick
AGN have been rescaled arbitrarily.
}
\label{fig:obsc_stack} 
\end{figure}

The fit to the stacked spectrum of unobscured sources
yields  a photon index of 1.71$^{+0.10}_{-0.06}$ 
and a reflection component of R=1.23$^{+1.12}_{-1.00}$.
 The uncertainties are large, however 
these results (which are reported in Tab.~\ref{tab:stack})  seem consistent with
the unified model which predicts a larger reflection
component for unobscured sources \citep[for a discussion
see e.g.][and references therein]{ajello08a}. 
In addition, our findings agree with the modeling 
of obscured sources in \cite{gilli07}, where the reflection efficiency 
for high inclination angles (expected for 
obscured AGN in the unified picture) is lower (0.88 rather than 1.3) than the one assumed for unobscured ones.
%
\subsubsection{The 15--195\,keV Spectrum of Compton-thick AGN}
 Finally we also investigate, for the first time, the average spectrum of Compton-thick AGN. Our first goal is to determine an empirical model which describes the 15-200\,keV  emission of Compton-thick AGN reasonably well and then later to interpret the features of the spectrum. Thus, we started fitting the stacked spectrum of the 9 Compton-thick AGN with a simple power-law model. The best-fit photon index is 2.04$\pm0.09$, but because of the spectral curvature this model represents a poor description of the data ($\chi^2/dof$=42.1/10). The fit improves ($\chi^2/dof$=16.1/9) if we use an absorbed power-law model. In this case the best-fit photon index and absorbing column density are respectively 2.48$^{+0.21}_{-0.18}$ and N$_H=4.7^{+2.4}_{-1.9}\times10^{24}$\,cm$^{-2}$. As a last step we tried to fit the stacked spectrum with an absorbed cut-off power-law model. We fixed the photon index of the power law to 1.8 to avoid degeneracy among the parameters. This model provides a good representation of the BAT data ($\chi^2/dof$=9.7/9).  The column density is consistent with being
Compton-thick (N$_{\rm H}=289^{+163}_{-131}\times 10^{22}$\,atoms cm$^{-2}$)
and the cut-off energy is 82$^{+39}_{-19}$\,keV. The results of this fit are summarized in Tab.~\ref{tab:stack}. 
However, we caution the reader this model (zphabs in \textit{Xspec})
takes into account only photoelectric absorption and it is used only
as a functional form to show that the average continuum 
of Compton-thick AGN is indeed very curved.
Fig.~\ref{fig:obsc_stack} shows the average spectra of unabsorbed,
absorbed and Compton-thick AGN and compares it to the general
shape of the Cosmic X-ray Background \citep{ajello08c}. 

The peak of the stacked spectrum of Compton-thick AGN (at z$\approx$0) is at almost twice the energy of the peak of the CXB (see Fig.~\ref{fig:obsc_stack}), testifying that if Compton-thick AGN are responsible for part of the emission at the peak of the CXB then the bulk of the population should be at z$\approx1$. This seems to be in agreement with the prediction of population synthesis models \cite[e.g.][]{gilli07,treister09}. 

\begin{figure}
\resizebox{\hsize}{!}{\includegraphics[angle=-90]{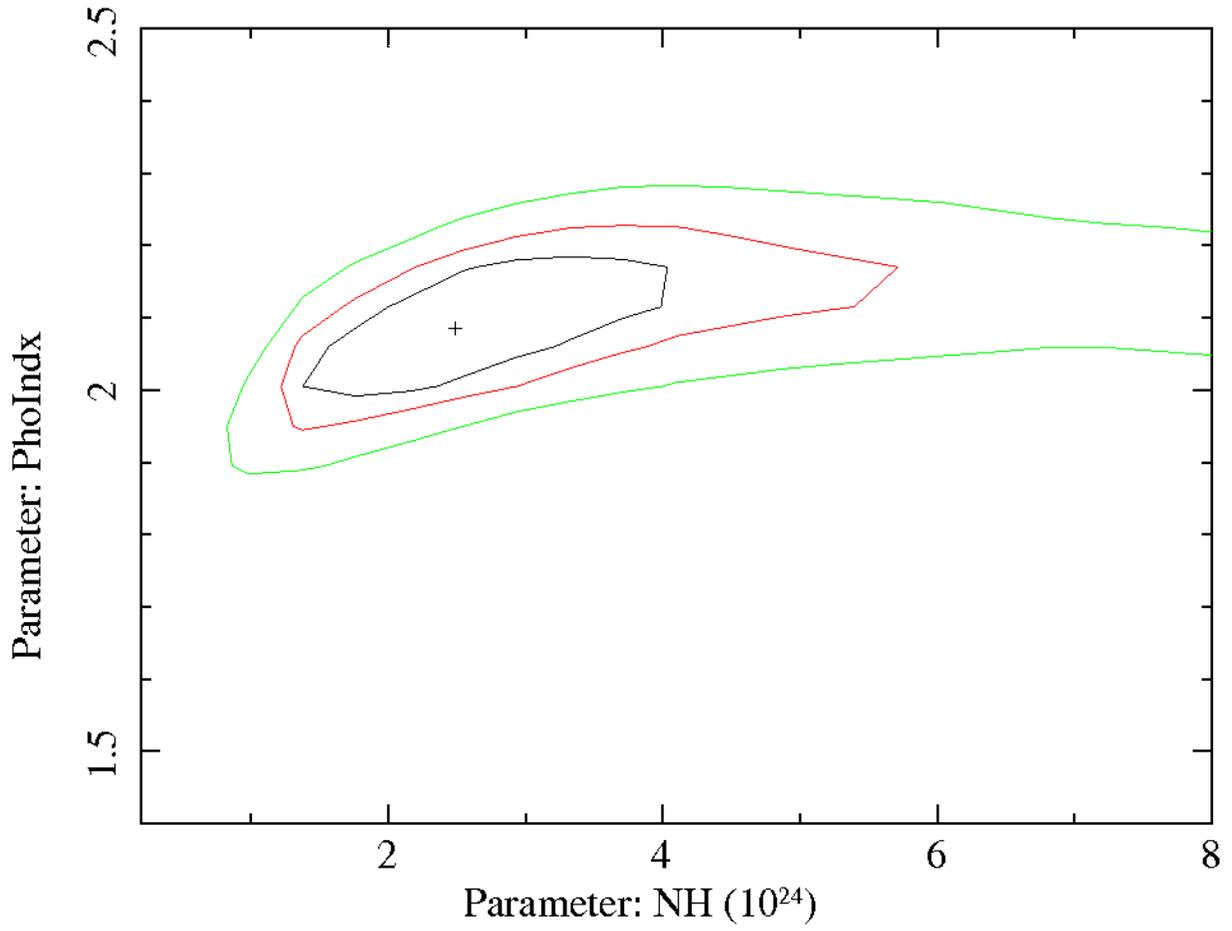} }
\caption{Confidence contours (1, 2, and 3\,$\sigma$) for the 
photon index and the absorbing column density of the MYTorus model
fitted simultaneously to all the 9 Compton-thick AGN spectra. } 
\label{cont} 
\end{figure}

 We also adopt a different strategy to check the results of the analysis 
of the stacked spectrum of the Compton-thick AGN. Instead of producing
a stacked spectrum, we performed a simultaneous fit to the 9 spectra.
In doing this we use the same baseline model for all the spectra
(e.g. a power law or an absorbed power law) with a free normalization
constant as to take into account the different source fluxes.
As a first test, we tried as before a fit with a simple power-law 
model. This fit yields a photon index of 1.94$\pm0.07$ 
(in reasonable agreement with what found before) and a $\chi^2/dof$=148.1/98.
We then tried fitting with an absorbed power-law model.
The improvement in the fit, with respect to the previous
one, is very large (e.g. $\chi^2/dof=$118.6/97) and the F-test confirms
that the probability of obtaining such improvement by chance is $\sim4\times10^{-6}$. The best-fit photon index is 2.29$^{+0.15}_{-0.14}$ and the 
column density is 2.4$^{+1.0}_{-0.8}\times10^{24}$\,cm$^{-2}$.
A fit with a cut-off absorbed power law model (with the photon index set to
1.8 as before) produces a slight improvement (e.g. $\chi^2=dof$=113.1/97) 
yielding an absorbing column density of 
1.8$^{+0.8}_{-0.3}\times10^{24}$\,cm$^{-2}$ and cut-off energy of 
128$^{+63}_{-34}$\,keV. Within the statistical uncertainties, the results
of this new analysis appear to be consistent with the results of the 
analysis of the stacked spectrum.

Our next step was then to fit the stacked spectrum of Compton-thick AGN with a more physical model. In this case we adopted an improved version of the model of  \cite{yaqoob97}
which fully treats relativistic Compton scattering, i.e. the MYTorus model by \cite{murphy09} and \cite{yaqoob10}\footnote{The model is available at www.mytorus.com}. This model provides tables for the attenuation of the continuum emission (transmitted through the torus) and the scattered component computed via Monte Carlo simulations \citep[a similar model can be also found in ][]{matt99}. In principle both the transmitted and the scattered components should be fitted to the spectrum to ensure self-consistency of the model. In practice, because of the many model parameters and the limited energy bandpass of BAT, a fit with both components was not successful (e.g. $\chi^2/dof>3$). However in this first exercise the normalization of the scattered component was a factor $>10$ larger then the transmitted one.  We then tried to fit the two components separately to understand whether one component is dominating over the other one. 
The best fit using the transmitted components yields a $\chi^2/dof$=20.6/9 and is still thus a fairly poor fit. Instead we achieved a good fit ($\chi^2/dof$=11.3/9) using the scattered component alone and an orientation of the torus (with respect our line of sight) of 60\,degrees. The best-fit photon index is 2.17$^{+0.10}_{-0.11}$ while the absorbing column density is N$_{H}=3.7^{+2.1}_{-1.9}\times10^{24}\,cm^{-2}$.
We also attempt a simultaneous fit with the MYTorus model to all the 9 
Compton-thick AGN spectra, leaving the normalization of each spectrum
to be a free parameter of the fit.
Again it appears that the scattered component
is dominating over the transmitted one. Indeed we achieve a good
fit to  the data using the scattered component alone (e.g. $\chi^2=$111.1/97)
and an inclination of the torus of $\sim$60\,degrees.
The best fit photon index is 2.08$\pm0.10$ and the absorbing column
density is 2.5$^{+1.8}_{-1.2}\times10^{24}$\,cm$^{-2}$.
Fig.~\ref{cont} shows the confidence contours of these two
parameters. 
If we remove the most obscured AGN from the fit (i.e. NGC 1068)
the index and N$_H$ become respectively 2.05$\pm0.10$ and 
 2.7$^{+2.2}_{-1.3}\times10^{24}$\,cm$^{-2}$ showing that our results
are not driven by just one particular source.

From the best fit using the MYTorus model (either to stacked spectrum
or the simultaneous fit) we derive that only $\sim$30\,\% of the intrinsic nuclear flux is observed in the 15--155\,keV band.
Finally, we note that the results presented in
this section do not change if we remove  the two brightest objects in the CT sample,  nor if we remove the most absorbed source (e.g. NGC 1068, see above) from the sample. Nevertheless, given the paucity of CT AGN in
our sample and the fact that they span one dex in absorbing column density,
the results of this section must be taken with care as they might turn out
not to be representative of the entire population of CT AGN.

\subsection{Luminosity distribution and spectral properties} 
\begin{figure}[ht]
\resizebox{\hsize}{!}{\includegraphics[]{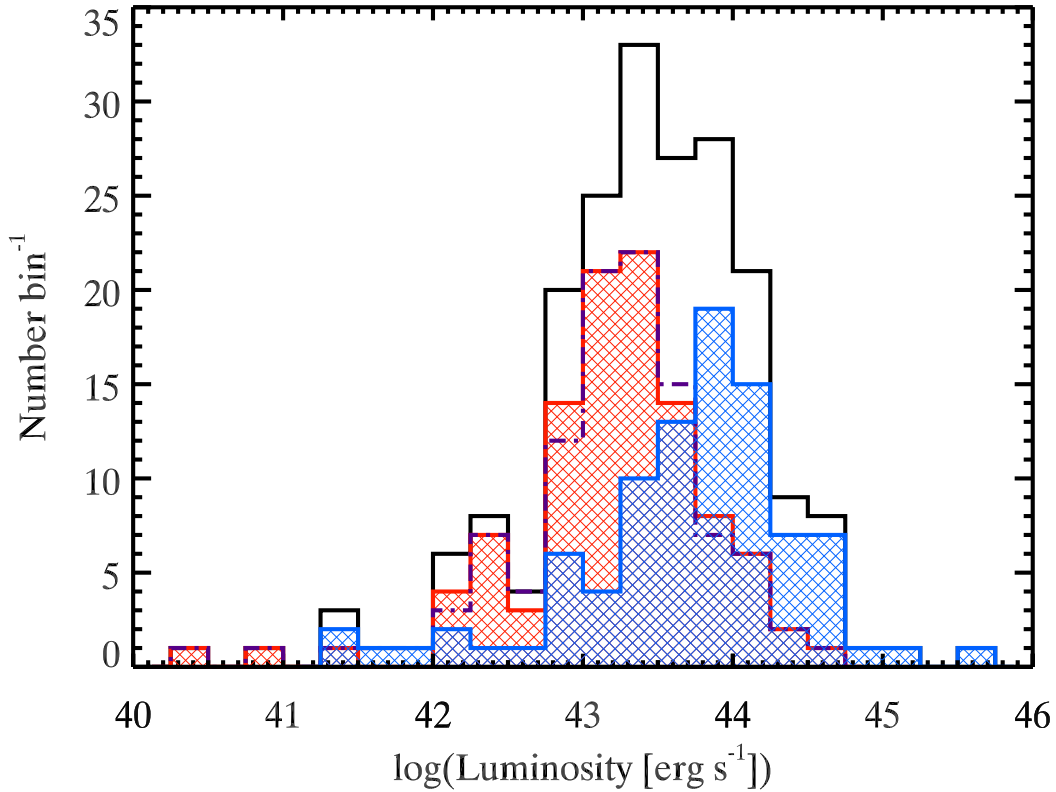} }
\caption{Luminosity distribution in the 15--55\,keV energy range for the complete catalogue (grey) and for the absorbed sources (i.e. log~\nh$\geq22$, 
red). We also show (blue) the distribution for the sources with log\nh\,$<22$.  The purple line shows the luminosity distribution of the absorbed sources without correction for absorption.  }   
\label{l_distr} 
\end{figure}

Luminosities of the Compton-thin AGN have been calculated through Eq.~\ref{lx}, therefore without taking absorption into consideration. We also tested whether the modeling of absorption introduced a bias in the distributions, even in the hard band sampled by BAT. To this aim we fitted all the sources with log\nh\ $>23.5$ taking Compton scatter into consideration (i.e. we used \texttt{cabs*zwabs*pow} in {\it Xspec}) and compared the resulting de-absorbed flux distribution with the one tabulated at the end of the paper. 
The \texttt{cabs} model has nonetheless some caveats that should be stated clearly: it assumes a constant Compton cross section equal to the Thomson one, so that it fails in describing the spectral hardening of the transmitted component due to the decay with energy of the Klein-Nishina cross section. Therefore is typically used for spectra in the 2--10 keV energy band. Moreover, it does not take into account scattering from material out of the line-of-sight.
The KS null probability (0.6) shows that indeed no appreciable difference is introduced by using a simple model for Compton-thin sources. As for the 9 Compton-thick objects, the fluxes have been de-absorbed as described in detail in \S4.2.
Fig.~\ref{l_distr} shows the distribution of luminosity in the energy range 15--55 keV 
for the complete sample of AGN along with the distributions for the 
absorbed and unabsorbed AGN. 
The median values of the two subclasses read 
 a logarithmic value of 43.2 and 43.8, respectively. 
A Kolmogorov-Smirnov test between these two populations reads a null hypothesis 
probability of $6.6\times 10^{-5}$, if we neglect those producing the peak at log\nh=20 in Fig.~\ref{nh_distr}. 
The distance between the two populations is exacerbated when taking into account all the sources with log\nh=20, since 
the KS probability drops to $\sim 10^{-9}$. Hence already by comparing the luminosity distributions there is evidence 
that the two populations are unlikely belonging to the same parent population, and this difference is independent from the modeling of absorption. 
This behavior is further discussed in detail in \S \ref{sec:lum}, where the X-ray 
luminosity functions of the two AGN classes are derived.

\begin{figure}
\resizebox{\hsize}{!}{\includegraphics[]{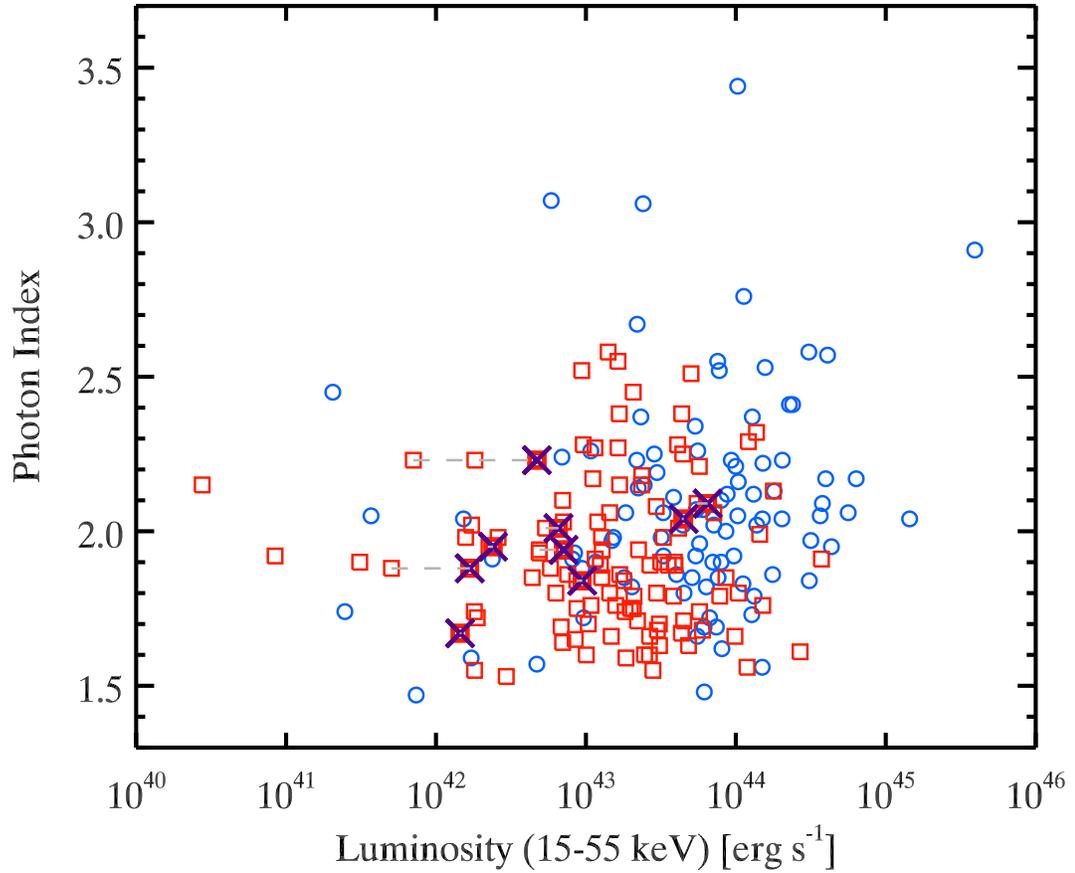} }
\caption{Photon spectral index  resulting from the simple power law fit as a 
function of the BAT luminosity in the 15--55 keV range. The unabsorbed sources are 
shown with blue circles, while the absorbed sources are shown in red squares. The 9 Compton-thick sources are highlighted with crosses, and their de-absorbed luminosity is connected to the absorbed one by horizontal lines. } \label{alfa_l}
\end{figure}

\emph{Photon index vs. Luminosity} --
In general, a correlation between the photon index and the accretion rate (expressed as the ratio of the bolometric luminosity and the 
Eddington luminosity) has been confirmed in numerous studies \cite[see][and references therein]{ishibashi10}. The 
general interpretation is that higher accretion rates lead to an increase in the photon density above the disk. This 
implies in general a more efficient cooling, and consequently steeper spectra. 
In this section we  test whether the hard X-ray power law index is correlated with the luminosity in the BAT range,  even if we are aware that the hard X-ray luminosity alone is an inaccurate proxy of the accretion rate. 
As it can be seen in Fig.~\ref{alfa_l} there is
 no indication that luminosity and photon index are correlated,  confirmed by the Spearman's rank coefficient and null hypothesis probability respectively equal to 0.19 and $\sim$0.01.
This might be produced by the fact that the objects in our sample have a broad distribution of black hole
masses (and Eddington ratios\footnote{By definition $\lambda_{Edd} \equiv L_{Bol}/L_{Edd}$ and $L_{Edd} \equiv 4\pi Gc M m_p/\sigma_T$, $G$ and 
$c$ being the gravitational constant and the speed of light, 
$M$ and $m_p$ the mass 
of the black hole and of the proton, $\sigma_T$ the Thompson cross section.}).
\cite{winter09} recently showed the absence (in the local BAT-selected AGN sample) on average, of this correlation 
between the 2--10 keV de-absorbed luminosities and $\Gamma$, as well as between a proxy 
%
of the Eddington ratio and the spectral index.  The correlation between the 2--10 keV photon index and the luminosity was found not to be significant in many papers \citep{reeves00,bianchi09}, and the significance is generally low also in the works where it is claimed as real \citep{dai04,saez08}.
It is worth noting that \cite{winter08} reported a positive correlation   
between the photon index and the 2--10 keV flux of  individual sources, i.e. at high fluxes the sources tend to show
steeper spectra.
Interestingly, \sw--BAT beamed AGN 
\cite[Fig. 2 in][]{ghisellini10,ajello09b} do show a rather remarkable relation between the 
15--55\,keV luminosity and the spectral index. This relation is even stronger when looking at samples selected in the \textit{Fermi}/LAT energy range. 
Nonetheless, the reason of the absence of this correlation in the objects of our sample is beyond the aims of this work.

\begin{figure}
\resizebox{\hsize}{!}{\includegraphics[]{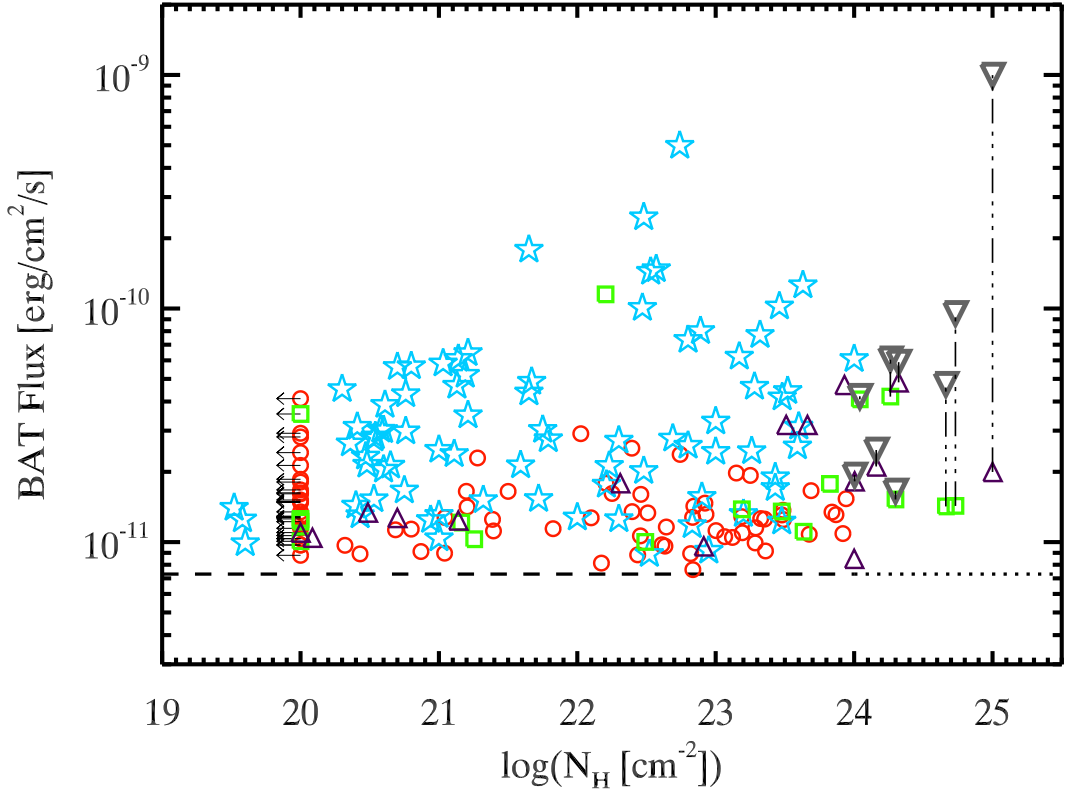} }
\caption{Scatter plot of the 15--55 keV fluxes of the sources  $vs.$ column density. Blue stars and purple triangles represent respectively the AGN already present in \cite{tueller08} and other references (see last table). Sources treated separately in this work are shown as red dots (follow up by \sw--XRT) and green squares (follow up by \xmm). The horizontal dashed line represents the sensitivity limit of the 3-year survey (note that we draw a dotted line in the CT regime, where also BAT
is biased, as discussed in \S~\ref{sec:ct}).  Absorbed and un-absorbed fluxes (connected by vertical dash-dotted lines) are drawn for the 9 Compton-thick sources. } 
\label{p_nh} 
\end{figure}

\emph{Hard X-ray flux versus Absorbing column density} --  Fig.~\ref{p_nh} shows
 the distribution of the sources in the flux--\nh\ plane. The absence of objects 
under $\sim 7 \times 10^{-12}$ erg cm$^{-2}$ s$^{-1}$  (which is represented by the horizontal
line) reflects the sensitivity of \sw--BAT in the energy range we 
selected. Indeed we see no correlation between these two parameters (Spearman's rank coefficient 0.04, and null hypothesis probability $P=0.64$).
Note that the absence of sources at small \nh\ and high fluxes,  i.e. the top left region of the plot, reflects the
tendency of the more luminous sources (which are intrinsically less numerous) of being unabsorbed.
 Indeed we showed in Fig.~\ref{lz} that at a higher flux the sources are typically  absorbed by columns in excess of $10^{22}$. This explains the handful bright objects, i.e.
with F$_X > 10^{-10}$ erg cm$^{-2}$ s$^{-1}$, in the range $22 <$log\nh$<23$. We discuss in further detail the nine Compton-thick objects in \S\ref{sec:ct}.
 
\section{Absorption in the local Universe}
\subsection{Observed \nh\ distribution} 
\begin{figure}
\resizebox{\hsize}{!}{\includegraphics[]{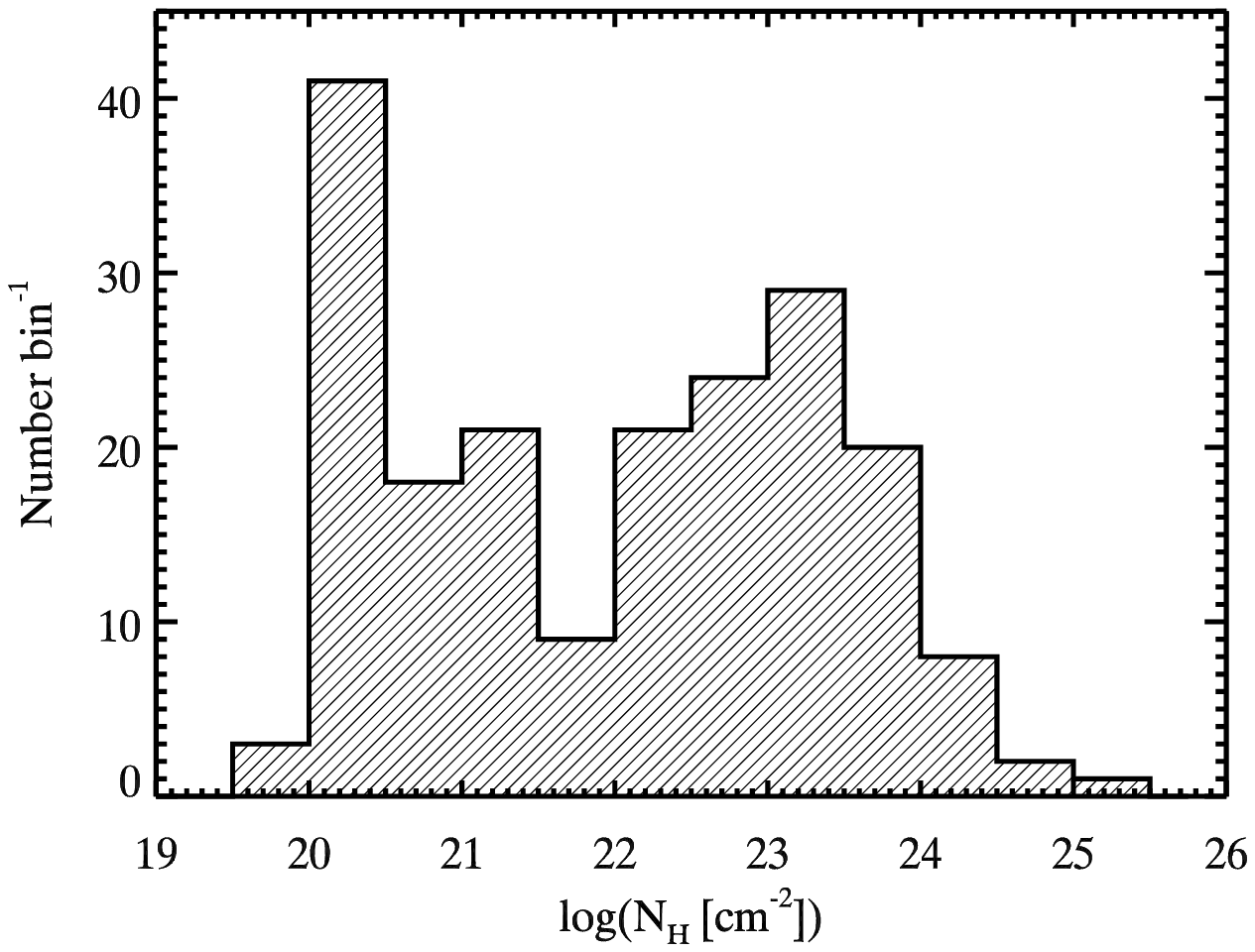} }
\caption{Observed column density distribution. The peak at log\nh= 20 is produced by the sources for which absorption was negligible (of the same order of the galactic one).} 
\label{nh_distr} 
\end{figure}

 We derived the photoelectric absorption by the combined fit of the BAT spectra (averaged over 3 yrs.) and the available follow up observations in the 0.3--10 keV energy range (e.g. \xmm\ or XRT). As described in \S2.3 we adopted the value reported in the literature for $\sim 50\%$ of the sample, and the specific reference is reported in the table at the end of the paper.
 The absorption distribution  is reported in Fig.~\ref{nh_distr}.
For 33 AGN (out of 197) the absorbing column density we measured  was found to be
consistent with (or smaller than) the 
Galactic absorption in the direction of the source. 
 When we could not constrain the absorption or when its value was consistent with the Galactic one, we put the value log\nh=20, which in turn produced the high peak in the distribution in the bin 20$<$log\nh $<$20.5.   
Note that there are a handful of cases taken from the literature in which the column density is found to be lower than $10^{20}$ cm$^{-2}$. 
 
When considering the whole AGN population  we find that 53$\pm4$\%
(1\,$\sigma$  statistical error) are absorbed by column densities $\geq$ 10$^{22}$ cm$^{-2}$.
We find that the number of objects whose \nh\ is greater then $10^{24}$ cm$^{-2}$ is 9/197. Thus the 
fraction of highly absorbed sources, know as Compton-thick AGN is $4.6 ^{+2.1}_{-1.5} \%$ (1\,$\sigma$  statistical error), all already known in the literature as extremely absorbed sources. 
These objects are highlighted  in the table at the end of the paper. 
The fraction of Compton-thick sources at $\sim 10^{-11}$\,erg cm$^{-2}$ s$^{-1}$ was predicted by population synthesis models to be,
at the typical fluxes sampled by BAT in the range 7--15\,\%  
\cite[see][and references therein]{comastri09,treister09}.
\\\\
An updated compilation of the most recent survey results
in the hard X-ray band (e.g. above 10\,keV) is reported in Tab.~\ref{tab:ctstory}. 
It is clear that most, if not all, of these results indicate a lack of Compton-thick AGN if compared to the $\sim$30\,\% 
fraction found by \cite{risaliti99}.

\begin{table}[ht]
\centering
\scriptsize
\caption{Observed fraction of absorbed and Compton-thick AGN, relative to the whole population for different X-ray surveys.
\label{tab:ctstory} }
\begin{tabular}{lcccccc}
\hline
\hline
Reference & \# Obj &Completeness & \% Absorbed   & \% C-thick  & Band & Instrument\\
        && & & & \scriptsize(keV)      \\
\hline
Markwardt et al., 2005		&  54	&95\,\%   & $\sim64$\%  & $\sim$10\% & 15--200 & \sw--BAT\\
Beckmann et al., 2006		& 36$^a$& 100\,\%  & $\sim64$\% & $\sim10\%$ & 20--40 & INTEGRAL\\
Bassani et al., 2006		& 56$^b$	& 77\,\%   & $\sim65\%$  &   $\sim14\%$ & 20--100 & INTEGRAL\\
Sazonov et al., 2007 		& 91   	& 93\,\%   & $\sim50\%$ & $\sim10$-$15\%$ & 17--60 & INTEGRAL\\
Ajello et al., 2008a 		& 24   	& 100\,\%  & $\sim55\%$  & $<$20\,\% & 14--170 & \sw--BAT\\
Tueller et al., 2008  		& 103  	& 100\,\%  & $\sim50\%$  &$\sim5\%$ & 14--195 & \sw--BAT \\
Paltani et al., 2008 		& 34$^c$& 100\,\%  & $\sim60\%$ & $<$24\,\%  & 20--60 & INTEGRAL\\
Della Ceca et al., 2008		& 62 	& 97\,\% & $\sim57\%$   & 0  &4.5--7.5 & \xmm	 \\
Malizia et al., 2009		& 79$^d$& 100\,\% & $\sim43\%$   & 7\%  & 20--40 & INTEGRAL\\
Beckmann et al., 2009		& 135$^e$&$\sim97$\,\% & $\sim44\%$   & $\sim$4\,\%  & 18--60 & INTEGRAL\\
\textbf{This work}			& 197  	& 100\,\% & $\sim53\%$   & $4.6 ^{+2.1}_{-1.5}$\,\%  & 15--195 & \sw--BAT \\ 
\hline
\end{tabular}
\begin{list}{}{}
\scriptsize


\item[$^{\mathrm{a}}$] The complete sample is 42 AGN, 36 of which are Seyfert galaxies.

\item[$^{\mathrm{b}}$] The complete sample is 62 AGN, 6 of which are Blazars and 14 are unindentified.
\item[$^{\mathrm{c}}$] Since the Paltani et al. sample may contain a fraction
of spurious sources, we restricted their sample to a limiting significance of
6\,$\sigma$. Above this threshold all sources are identified 
(see Tab.~2 in \citealp{paltani08}).

\item[$^{\mathrm{d}}$] There are 88 objects reported to be at significance $>5.2 \sigma$. 79 of those are Seyfert galaxies, the remaining being Blazars.

\item[$^{\mathrm{e}}$] The complete sample comprises 187 objects with $>3\sigma$ significance in the 18--60 keV energy band. According to the authors \cite[see Sect 4.1 in][]{beckmann09} there are 135 Seyfert galaxies with measured absorption. Only 7/187 sources are listed generically as AGN without information on redshift.

\end{list}
\end{table}

\subsection{The BAT bias and the intrinsic \nh\ distribution}\label{sec:ct}
\begin{figure}[h!]
\resizebox{\hsize}{!}{	\includegraphics{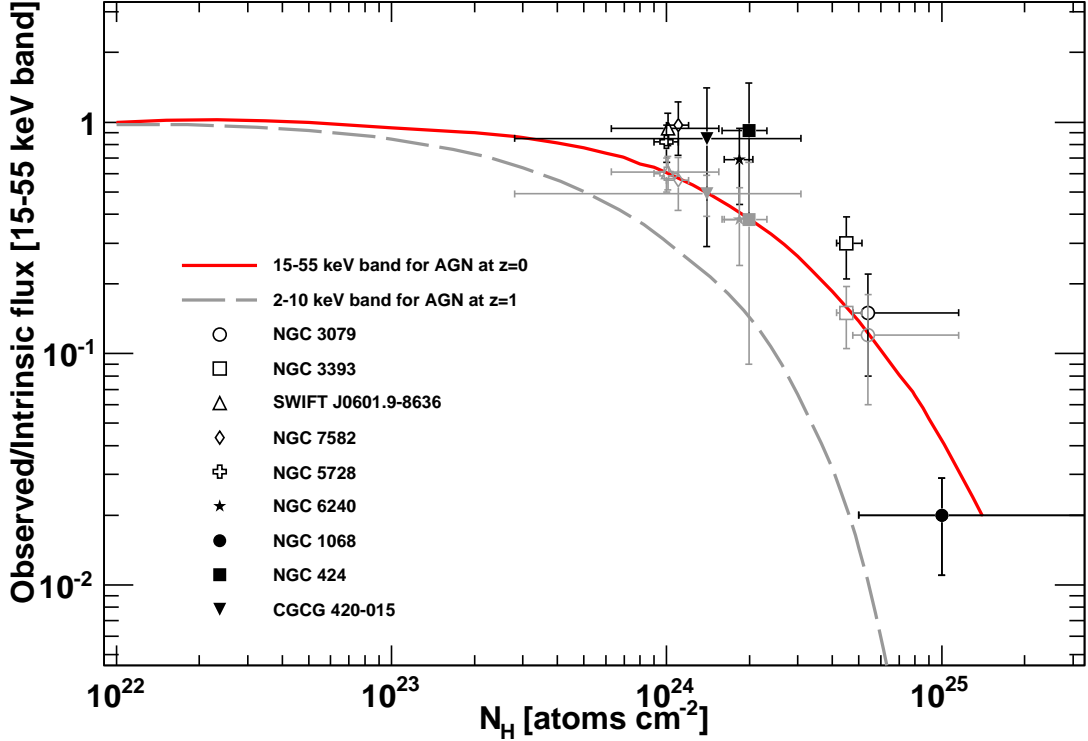} }
	\caption{Ratio of observed to intrinsic nuclear flux for an AGN, with a power-law spectrum
with index of 1.9, as a function
of the column density of the torus as seen edge-on. 
The red solid line shows the ratio for the 15-55\,keV
band for an AGN in the local Universe 
(e.g. the typical case for the BAT survey). The dashed line shows the ratio 
for the 2-10\,keV band for an AGN at redshift $\sim1$, i.e. the typical 
redshift for AGN detected in the \xmm/Chandra surveys \cite[see e.g.][]{akylas06}.
Note that the BAT survey is unbiased (e.g. ratio $\approx$1) for
logN$_{\rm H}\leq$23.5.  The black data points show the ratio of observed to
intrinsic flux estimated for the BAT Compton-thick AGN taking
into account only photoelectric absorption. The gray data points show
the ratio also when Compton scattering is fully taken into account.
	\label{fig:nh_intr}}
\end{figure}

BAT is the least biased X-ray instrument, particularly when
comparing it to 2--10\,keV telescopes, for the detection of obscured
objects. Nonetheless, even in the $>10$\,keV band a relevant
fraction of the source flux might be lost if the source is Compton-thick. 
In order to show this effect we computed the ratio between the 
observed and the intrinsic
nuclear flux of an AGN for increasing column densities. In this
exercise, we took both photoelectric absorption and Compton scattering
into account using  MYTorus. 
The nuclear, intrinsic, emission has been modeled as a power law with a photon index of 1.9.
The results are shown in Fig.~\ref{fig:nh_intr}. From this plot it is
apparent that the BAT survey is unbiased up to logN$_{\rm H}\approx$24
and then becomes biased against the detection of Compton-thick objects.
Furthermore, the absorption bias affects much more severely the 
2-10\,keV band already for logN$_{\rm H}\geq$23 
(see same  Fig.~\ref{fig:nh_intr}).
  
We also performed joint spectral fits of \xmm\ and \sw--BAT
data for all the Compton-thick objects in our sample to determine
the fraction of the intrinsic flux which is seen in the 15-55\,keV band.
The details of this analysis will be reported in a future publication
(Burlon et al. in preparation), however
the results are summarized in Tab.~\ref{tab:cthick} 
and  in Fig.~\ref{fig:nh_intr}. Only for SWIFT~J0601.9-8636,
NGC~5728, CGCG~420-015, and NGC~1068 we used the values available in the literature.
Indeed, \xmm\ data are not available for SWIFT~J0601.9-8636 and NGC 5728, so 
we used the values derived with Suzaku by \cite{ueda07} and \cite{comastri10}, respectively. 
 NGC~7582 was largely discussed in \cite{piconcelli07} and \cite{bianchi09b}, as for the variations of a factor $\sim2$ in the thicker absorber. We used the \xmm\ observation taken in 2005, and therefore tagged this object as Compton-Thick.
Variations of the thick absorber can in principle take place for other sources as well, altering the fraction of CT objects according to the observation used for the analysis.
The Compton-thick nature of CGCG~420-015 (aka IRAS~04507+0358)  was discovered by \cite{severgnini10}. It was known to be a highly absorbed Sy2 galaxy, but with the use of deep (100 ks) Suzaku observations, in addition to the ones available, the authors found \nh\ to be $\sim1.3-1.5\times 10^{24}$ cm$^{-2}$, with minor variations according to the model used to fit the combined spectra.
NGC 1068 is a complex object which has been analyzed in detail
in the past. According to \cite{matt04}, the nuclear emission is
completely suppressed and the source is seen only in reflected light.
The column density is probably in excess of 10$^{25}$\,atoms cm$^{-2}$
\citep{matt04} and the reflection component is of the order of a {\it few} \% of the nuclear
flux \citep{iwasawa97}.
For these reasons the position on the plot of NGC 1068 should be considered
tentative.   Fig.~\ref{fig:nh_intr} also shows the difference
in the observed-to-intrinsic flux ratio when also Compton scattering
is taken into account (using the MYTorus model). 
Note however that modeling with MYTorus contains an implicit assumption, i.e. that the scattering material has a toroidal geometry with given parameters (half-opening angle of 60$^{\circ}$, corresponding to a covering factor of $\Delta \Omega/4\pi = 0.5)$. The line-of-sight of the observer is a fixed parameter of the fit.

It is interesting to note that there is relatively good
agreement between the model line and the observations of single Compton-thick
objects detected in the \sw--BAT survey.
In Fig.~\ref{p_nh} we showed the fluxes of the 9 Compton-thick AGN, before and after 
the correction for the missed flux, connected by dash-dotted lines. The horizontal line
representing the limiting flux of the survey was turned into a dotted line to visually 
indicate the increasing bias in the CT regime.
\begin{table}[ht]
\centering
\scriptsize
\caption{Ratio of the observed flux with respect to the nuclear (unabsorbed) flux in the 15--55 keV band for the 9 Compton-thick objects in the sample. Errors are quoted at 90\,\% CL. Four sources were fitted using \xmm\ and \sw--BAT. For the remaining objects we used values from the literature.}
\smallskip
 \label{tab:cthick}
\begin{tabular}{lcccc}

\hline
\hline
Source  & 	F$_{\rm{obs}}$/F$_{\rm{nucl}}$  	& Error 	& \nh\ 	& \nh\ error  \\
        &             									&	    &  (10$^{24}$\,cm$^{-2}$) & (10$^{24}$\,cm$^{-2}$)  \\
\hline

NGC 3079      &     0.15 &  0.07 &      5.40  &    (-0.65,6.10)\\

NGC 3393     &      0.30 &  0.10  &     4.50  &    (-0.36,0.62)\\

SWIFT J0601.9-8636 $^a$ & 0.94	 & 0.15	&      1.01	&      (-0.38,0.54)	\\

NGC7582       &     0.97 &  0.25   &   	1.10	 &   (-0.05,0.05)\\

NGC 5728 $^b$      &     0.82 &  0.15    &   1.0  &    (-0.1,0.1)\\
 
NGC 6240     &      0.69&   0.25 &      1.83 &   (-0.23,0.22)\\

NGC 1068 $^c$     &      0.02 &  0.01&       $>$10  &   (-5, $>$10) \\

NGC 424 &	   0.92&	 0.56	&      1.99&		 (-0.40,0.32) \\

CGCG 420-015 $^d$ &	0.85	&	$\sim$0.20	&	1.46 &	(-0.11,0.07)\\
\hline

\end{tabular}

$^{\mathrm{a}}$ \cite{ueda07}
$^{\mathrm{b}}$ \cite{comastri10}
$^{\mathrm{c}}$ \cite{matt04} and \cite{iwasawa97}
$^{\mathrm{d}}$ \cite{severgnini10}
%

\end{table}

Essentially the absorption bias limits the detection of Compton-thick
objects only to those with bright (intrinsic) fluxes and 
in the very local Universe.
Indeed three of the most famous Compton-thick objects
(NGC 1068, NGC 4945 and the Circinus galaxy) are also  among the closest
known AGN.
Thus the distribution reported in Fig.~\ref{nh_distr}
compares sources detected at different limiting {\it intrinsic}  fluxes.
It is possible to correct for this 
effect by taking into account the
selection effect due to the large column density.
The {\it intrinsic} absorption distribution can be expressed as:
\begin{equation}
\frac{dN}{dLogN_{\rm H}}=\int^{S_{max}}_{S_{min}} \frac{dN}{dS}(N_{\rm H})\ dS ,
\label{eq:nh_intr}
\end{equation}

where the d$N$/dLog\nh\ is in unit of sr$^{-1}$ per logarithmic
bin of \nh\ , $S$ is the observed source flux, 
 and d$N$/d$S$(\nh) is the log$N$--log$S$ of sources
in a given log\nh\ bin. The minimum {\it observed} flux ($S_{min}$)
of integration should be set so that the limiting {\it intrinsic}
flux is the same for all the bins.
In this way the absorption distribution derived is
representative of the density of sources at the same limiting intrinsic
flux.

The relationship between observed and intrinsic flux can
be expressed as $S^{obs}$ = $K(N_{\rm H})\ S^{intr}$  where
$K(N_{\rm H})$ is the ratio plotted in Fig.~\ref{fig:nh_intr}.
Thus $S_{min}$ can be set to 10$^{-11}\times K(N_{\rm H})$
to produce a uniform absorption distribution for sources with
an intrinsic flux greater than 10$^{-11}$\,erg cm$^{-2}$ s$^{-1}$.
Eq.~\ref{eq:nh_intr} can thus be rewritten as:

\begin{equation}
\frac{dN}{dLogN_{\rm H}}= \frac{A(N_{H})}{1-\alpha}\left[ S^{1-\alpha}_{max} 
- (10^{-11}K(N_{\rm H}))^{1-\alpha} \right]\ \ ,
\label{eq:nh_intr2}
\end{equation}
where $A(N_{\rm H})$ and $\alpha$ are the normalization and the index
of the $log$ N--$log$ S in a given logN$_{\rm H}$ bin.
Here we assumed that the source count distribution can be approximated
with a power-law function (e.g. $dN/dS=A S^{-\alpha}$). This assumption
is well justified by the fact that the source count distribution
of the  entire BAT sample is well represented by a single power law 
\citep[e.g. see ][]{ajello09b} and that the source count distribution
of AGN shows a break at much lower fluxes \citep[e.g.][]{cappelluti07}.
\begin{figure}[h!]
\resizebox{\hsize}{!}{\includegraphics[scale=0.6]{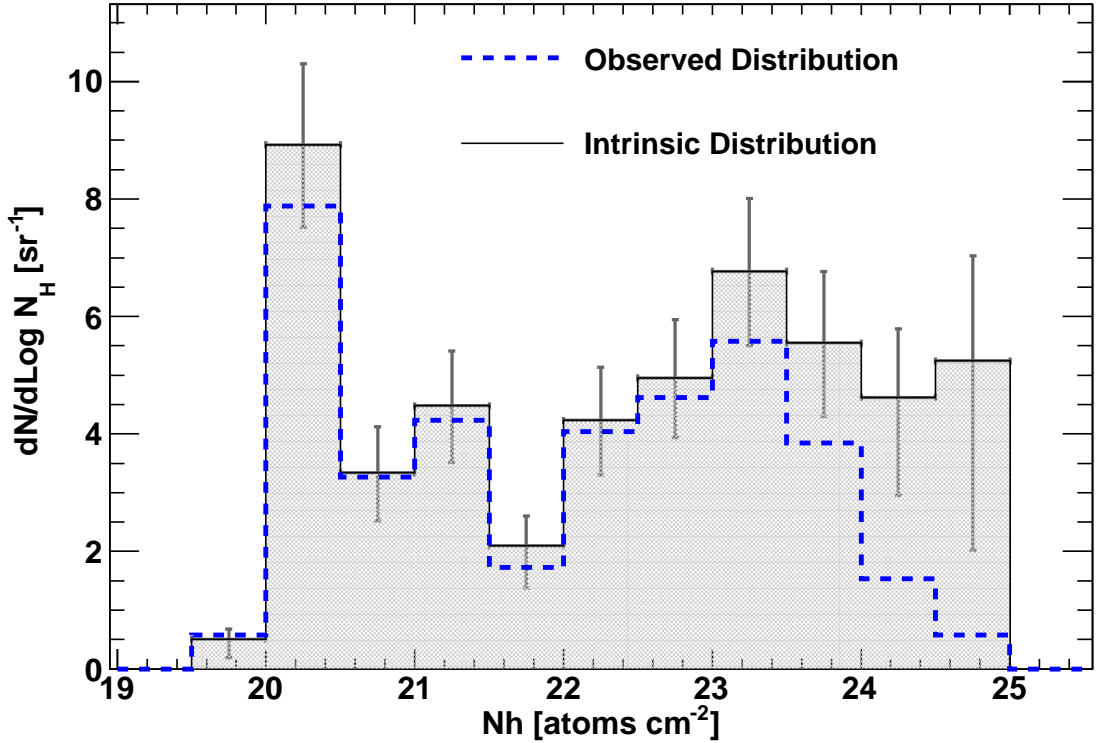} }
	\caption{Intrinsic absorption distribution compared to the observed
one (dashed line). Errors bars are derived from counting statistics in
each bin.
	\label{fig:nh_intr2}}
\end{figure}
In order to derive the intrinsic absorption distribution we used
0.25 dex logN$_{\rm H}$ bins. For each bin, a source
count distribution has been constructed and a power law has been
fitted to it employing a Maximum Likelihood algorithm. This power law
was then used to estimate the density of objects (per steradian) in
a given log\nh\ bin (i.e. Eq.~\ref{eq:nh_intr2}). 
All the  power-law indices were found to be compatible with a Euclidean
index (i.e. 2.5 for a differential distribution). The index $\alpha$ has
been fixed to 2.5 for all those bins with less than 5 objects.
We checked that fixing the index to 2.5 for all bins or allowing it to vary
does not change the results presented here. For each logN$_{\rm H}$ bin
the error on the density of sources is derived from the Poisson error
on the number of sources present in that bin to preserve the original 
counting statistics.

The intrinsic absorption distribution is shown in Fig.~\ref{fig:nh_intr2}.
From this we derive that the {\it intrinsic} 
fraction of Compton-thick sources (logN$_{H}>$24)
is 20$^{+9}_{-6}$\,\%, where the error is only statistical.
It is apparent that the observed distribution starts to deviate from
the observed one only for logN$_H\geq$23.5. We also derive that
the  {\it intrinsic} fraction of absorbed sources is 65$\pm4$\,\%.
A systematic error in estimating the {\it intrinsic} fraction 
of Compton-thick sources might arise from both the power-law indices
fitted in Eq.~\ref{eq:nh_intr2} and from the $K(N_{\rm H}$) correction
factor (e.g.  the curve plotted in  Fig.~\ref{fig:nh_intr}).
Fixing all the $\alpha$ parameters to 2.5 or allowing all of  them to vary
changes the above fraction of about 1\,\%. Thus the exact shape
 of the log$N$--log$S$ in each logN$_{\rm H}$ bin does not contribute
a large systematic error.

On the other hand, the knowledge of the fraction of transmitted 
flux (e.g. $K(N_{\rm H}$) and Fig.~\ref{fig:nh_intr}) plays a major
role in the derivation of the density of Compton-thick AGN.
The angle at which our line-of-sight intersects the torus
and the power-law index of the intrinsic AGN spectrum can modify the
fraction of Compton-thick AGN. Playing with these different parameters
we derive that the systematic uncertainty on the fraction of Compton-thick
AGN is $\sim$5\,\%.

Also \cite{malizia09}, using INTEGRAL, showed that the fraction
of Compton-thick AGN is likely larger than the observed $\sim$5\,\%.
Instead of correcting for the missing population, they adopt a
redshift cut (z$\leq$0.015) which would ensure, according to the
authors, to have a complete sample. In their sample of 25 AGN, they
found 6 Compton-thick AGN, thus the fraction of Compton-thick AGN
is 24$^{+11}_{-9}$\,\% of the total population in agreement with our estimate.

 As visible from Fig.~\ref{fig:nh_intr}, there is also 
a slight overestimate (although compatible within 1\,$\sigma$ with
the observed density) of the intrinsic density of objects with LogN$_H\approx20$.
This is due to the fact that a few objects in that bin have actually
a lower column density that could not be effectively constrained in the 
0.3--10\,keV energy band. For all those objects the  LogN$_H\sim20$ can be
considered an upper limit to the true absorbing column density.
Because of this, the source count distribution in the  LogN$_H\approx20$ bin
tends to overestimate the true intrinsic density. However, as 
seen in the Fig.~\ref{fig:nh_intr} this effect is small.

\section{Anti-correlation of Absorption and Luminosity}

According to the AGN unified model \citep{antonucci93, urry95} all the different properties of AGN can be ascribed 
solely to orientation effect. Thus one should not observe variations of any other
property with e.g. luminosity, accretion rate and redshift.
However, already 30 years ago, \cite{lawrence82} reported the discovery
of the anti-correlation of the fraction ($F_{\%}$) of obscured AGN (relative to the whole population)
and luminosity.
More recently different authors addressed the same issues with contradicting 
results. For example some studies \cite[e.g.][]{treisterurry06,lafranca05,dellaceca08,winter09,brusa10} suggest that $F_{\%}$ decreases with X-ray
luminosity, while \cite{dwelly06} point to $F_{\%}$ being independent of L$_X$.
Also \cite{sazonov04}, \cite{sazonov07}, and \cite{beckmann09} pointed out 
that an anti-correlation of the fraction 
of absorbed sources and X-ray luminosity seems to exist. 

\begin{figure}
 \centering
 	\subfigure[
The red dashed line represent the fit with Eq.~\ref{gillilike}. Grey stars represent, for comparison, data from \cite{beckmann09}. Luminosities measured by INTEGRAL have been converted into the 15--55 keV energy range. ]{\includegraphics[width=9cm,height=6cm]{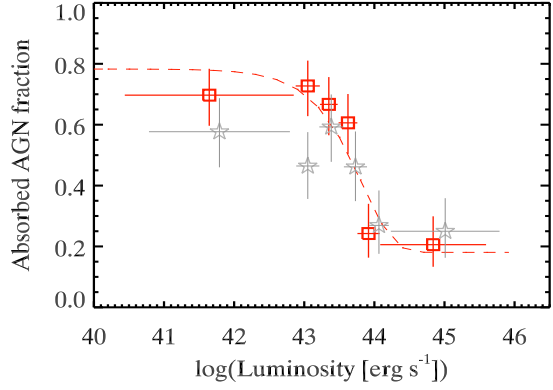}}
	\subfigure[Same as above (a) but with a cut in luminosity, as discussed in the text. The dashed (dotted) line represents a linear fit to the data ($1\sigma$ uncertainties). ]
	{\includegraphics[width=9cm,height=6cm]{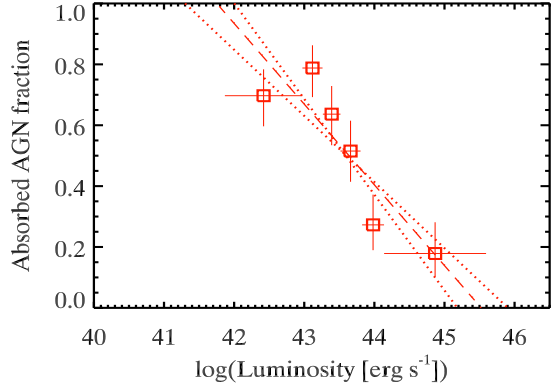}}
 \caption{
Fraction of the obscured (i.e. \nh\ $\geq 10^{22}$ cm$^{-2}$) AGN
as a function of 15--55\,keV luminosity (red squares).
The data have been grouped such as to have an equal number of sources per luminosity bin. Errors on fraction of absorbed AGN have been computed using Binomial statistics \cite[see][]{gehrels86} and are drawn at 1\,$\sigma$ level.}
 \label{anticorr}
 \end{figure}

Fig.~\ref{anticorr} shows how the fraction of obscured AGN (those with logN$_{\rm H}\geq22$)
changes as a function
of X-ray luminosity (in the 15--55\,keV band) in our survey.
The width of each bin has been chosen so that the number of object per bin is constant ($\sim33$). 
The  errors on the number of absorbed sources and the total number of sources per bin have been propagated with the 
recipes for Binomial statistics \cite[see][in particular Tab.~6]{gehrels86}. 
Binomial statistics apply specifically when dealing with ratios of small numbers.
 The data are correlated, as the Spearman's rank correlation coefficient is r$_s$ = -0.94. 
The probability of a chance correlation ($4.8\times 10^{-3}$) shows that the correlation is true at the $2.9\sigma$ confidence level. 

We also show for comparison the INTEGRAL results of \cite{beckmann09}, by converting
the INTEGRAL 20--100\,keV luminosities into 15--55\,keV luminosities and selecting
only non-blazar sources with z$<$0.3. The two data-sets are in reasonable agreement
although the lower fraction of absorbed AGN in the INTEGRAL sample (with respect to
the BAT sample) might be ascribed to completeness issues of the former.

In the BAT sample, the absorbed AGN fraction displays
a flattening in the low luminosity regime, and the correlation becomes
 clearly non-linear ($\chi^2/dof~\gsim~3.5$).
Thus, we tried to fit the fraction of absorbed AGN
with the empirical function proposed 
by \cite{gilli07} of the form:
\begin{equation} \label{gillilike}
F_{\%}(L_X) = R_{low} e^{(-L_X/L_C)} + R_{high} [1-e^{(-L_X/L_C)}]
\end{equation}
where R$_{low}$ is the low luminosity asymptotic behavior, R$_{high}$ the high luminosity one, and $L_C$ is a 
``critical'' luminosity at which the drop occurs. We fitted this function to the data: the best fit values for 
R$_{low}$, R$_{high}$, and log(L$_C$) are respectively 0.8, 0.2, and 43.7. 
This fit yields a $\chi^2/dof=1.3$ which shows that this fit is better than a simple linear
relation between the obscured AGN fraction and (the logarithm of) luminosity.
This result will be  discussed in details in in \S~\ref{sec:lum}.

In the literature \cite[e.g.][]{hasinger08,dellaceca08}  the minimum luminosity 
considered for an AGN in the 2--10 keV  energy range, was typically greater than 10$^{42}$  erg s$^{-1}$. 
Since the low luminosity tail of our distribution (see Fig. \ref{l_distr}) extends to $\sim 2.8 \times 10^
{40}$ erg s$^{-1}$ we checked, for the sake of completeness, if by introducing a cut at 10$^{42}$  erg s$^{-1}$, changed our findings. 
When transforming the 2--10\,keV luminosity of 10$^{42}$\,erg s$^{-1}$ to the 15--55\,keV
band, we find that only 7 sources fall below this limiting luminosity.
The level of correlation of the data remains unchanged, 
while instead the $\chi^2/dof$ value for a linear fit, decreases to $\sim1.8$.
These results are shown in Fig.~\ref{anticorr} (b). It is still apparent
that the linear fit (which shows a slope of  $-0.26\pm0.05$)
is not a good representation of the data since the first bin 
and the last two show a flattening of the fraction of absorbed AGN.
Finally, it is worth considering that (i) the contribution from the stellar population 
hardly  extends above $\sim$10$^{41}$\,erg s$^{-1}$ \cite[][]{ranalli03,norman04}, and (ii) in
the 15--55\,keV band this contribution is expected 
to decrease to even a lower fraction of the ``bolometric'' luminosity \cite[see e.g.][]{voss10}.

\section{Luminosity Functions of AGN} 
\label{sec:lum}

\begin{figure}
\resizebox{\hsize}{!}{\includegraphics[]{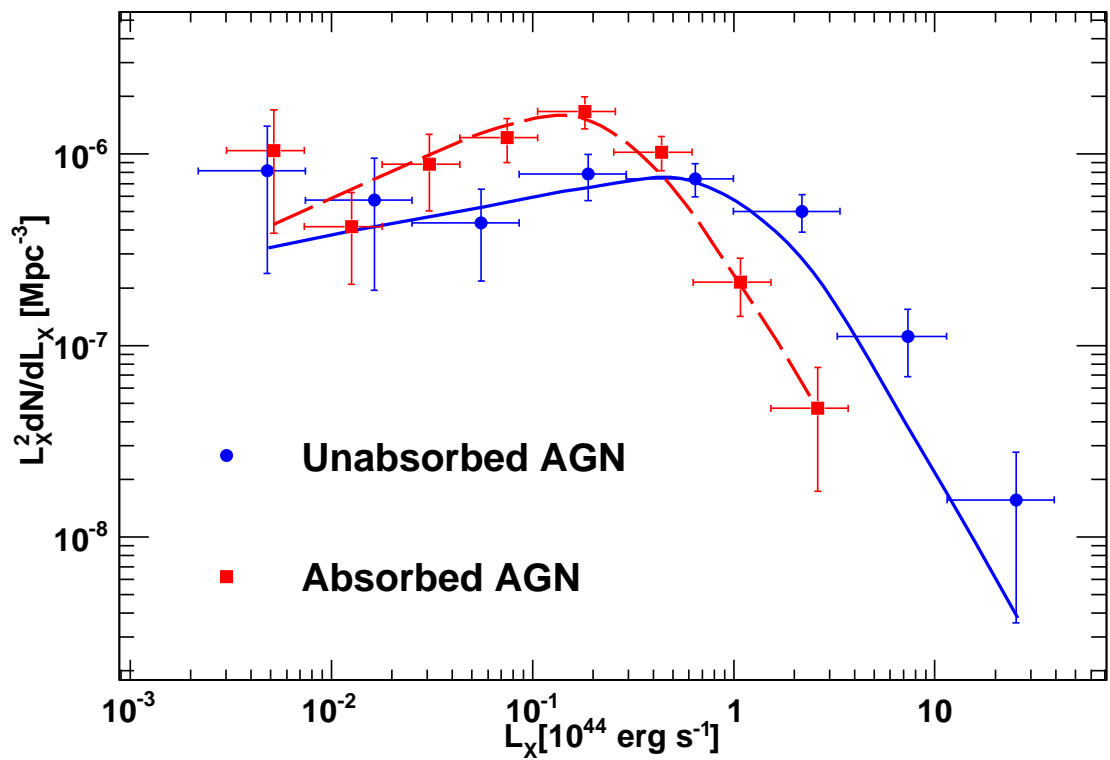} }
\caption{Luminosity function of absorbed (logN$_H\geq$22, squares)
and unabsorbed AGN (logN$_H<$22, circles) with the respective
best-fit models (solid and dashed lines). The luminosity functions
have been multiplied by $L_{X}^2$ to highlight the different positions
of the break luminosities.
\label{fig:xlf} }
\end{figure}

We estimate the X-ray luminosity function (XLF) of  AGN
using the  $1/V_{MAX}$ method (equivalent in our
formalism to the $1/V_{a}$ method). In this framework the
XLF can be expressed as:
\begin{equation}
\Phi(L_{X},z) = \frac{dN}{dL_{X}} = \frac{1}{\Delta\ L_{X}}\
\sum_{i=1}^{N} \frac{1}{V_{MAX,i}}
\end{equation}
where $V_{MAX,i}$ is the maximum comoving volume associated with the 
$i_{th}$ source. This is defined as:
\begin{equation}
V = \int_{z=0}^{z=z_{MAX}} \frac{dV}{dz}\ \Omega(L_i,z) dz,
\label{eq:vvmax}
\end{equation}
where $dV/dz$ is the comoving volume element per unit redshift and 
unit solid angle \citep[see e.g.][]{hogg99}, $z_{MAX}$ is the maximum
redshift after which the sources would not be detected anymore
in the BAT survey
 and $\Omega(L_i,z)$
is the sky coverage of the survey.
The XLF of the two different classes of AGN, obscured and unobscured,
is reported in Fig.~\ref{fig:xlf}.

We made a Maximum Likelihood fit to the two different datasets
using a 
broken power-law of the form \citep[see e.g.][]{ueda03,hasinger05}:
\begin{equation}
\Phi(L_X,z=0) = \frac{dN}{dL_X}=
\frac{A}{ln(10)L_X}\left[ \left(\frac{L_X}{L_*} \right) ^{\gamma_1} 
+ \left(\frac{L_X}{L_*}\right)^{\gamma_2} \right]^{-1}
\label{eq:2pow}
\end{equation}

The ML estimator can be expressed as:
\begin{equation}\label{eq:ml_est}
\mathcal{L}= -2\sum_i {\rm ln} \frac{ \Phi(L_{X,i},z_i) V(L_{X,i},z_i)}
{\int  \Phi(L_{X},z)  V(L_{X},z) dL_{X}}\ .
\end{equation}

\begin{table}[ht]
\centering
\scriptsize
\caption{Best-fit parameters of X-ray Luminosity Functions in the 15-55\,keV band\label{tab:xlf_fits}}
\begin{tabular}{lccccc}
\hline
\hline

SAMPLE           & \# Objects       &
Norm.$^1$            &  L$^{*2}$           &
$\gamma_1$       &  $\gamma_2$ \\

\hline
ALL       & 199 & 1.53e-5 & 0.53$^{+0.15}_{-0.15}$ & 0.74$^{+0.07}_{-0.08}$ & 2.60$^{+0.19}_{-0.20}$\\

ABSORBED  & 105 & 2.59e-5 & 0.26$^{+0.08}_{-0.07}$ & 0.58$^{+0.12}_{-0.13}$ & 2.75$^{+0.34}_{-0.30}$\\
ABSORBED$^3$  & 99  & 3.93e-5 & 0.26$^{+0.14}_{-0.09}$ & 0.51$^{+0.28}_{-0.34}$ & 2.63$^{+0.38}_{-0.31}$\\
UNABSORBED & 92 & 1.90e-6 & 1.34$^{+0.48}_{-0.38}$ & 0.80$^{+0.11}_{-0.12}$ & 2.88$^{+0.37}_{-0.31}$\\

\hline
\end{tabular}
\begin{list}{}{}
\item[$^{\mathrm{1}}$] Normalization of the XLF expressed in
units of erg$^{-1}$\,s\,Mpc$^{-3}$.
\item[$^{\mathrm{2}}$] In units of 10$^{44}$\,erg s$^{-1}$.
\item[$^{\mathrm{3}}$]  {\it Intrinsic} XLF of absorbed AGN. 
The luminosity of the absorbed AGN with LogN$_{H}\geq23.5$ have been
de-absorbed with the method described in $\S~$\ref{sec:ct}.

\end{list}
\end{table}

The best-fit parameters are obtained by minimizing $\mathcal{L}$.
Their  1\,$\sigma$ error 
are   computed by varying the parameter
of interest, while the others are allowed to float, until an increment
of  $\Delta \mathcal{L}$=1 is achieved. This gives an estimate of the 68\,\%
confidence region for the parameter of interest \citep{avni76}.
The likelihood function does not depend on the normalization A since
it cancels out in Eq.~\ref{eq:ml_est}.
Once the slope $\alpha$ is determined, the normalization is derived 
as the value which reproduces the number of observed sources.
An estimate of its statistical error is given by the Poisson error on the number
of sources used to build the XLF.

The results of the ML fits to the XLFs of whole population of AGN and 
 obscured and unobscured subclasses are summarized in Tab.~\ref{tab:xlf_fits}.
 We focussed mainly on the difference between the two subsamples 
 of absorbed and unabsorbed AGN, but we used the total XLF in Fig.~\ref{fig:xlf_boot_1000}
 in order to account for the obscuration-luminosity relation. 
It is apparent that the XLFs of the two classes of objects are not
the same. In particular the 'break' luminosity $L^*$ is different
at $\sim$2.8\,$\sigma$ level, with absorbed AGN having on average
lower luminosity than unabsorbed ones. Also \cite{dellaceca08}, 
analyzing a small sample of XMM-Newton AGN, found different XLFs
for obscured and unobscured sources. However, in their case
they cannot allow (presumably due to the low number of sources)
L$^*$ to be a free parameter  of the fit. In our case, this can be
 done and there is evidence (albeit marginal) that the typical luminosity
of absorbed and unabsorbed AGN is different.
The difference between the luminosity functions of absorbed and unabsorbed
objects is however not a surprise. Indeed it is expected in
view of the anti-correlation of the absorption fraction and
luminosity (e.g. Fig.~\ref{anticorr}).
The two luminosity function are equal at a luminosity of 
4$\times10^{43}$\,erg s$^{-1}$. This is exactly the luminosity
at which the fraction of absorbed objects is 0.5 (see Fig.~\ref{anticorr}). 
For the very first time this trend is clearly seen in the 
luminosity function of absorbed and unabsorbed objects (as
derived from the same energy band).

 \begin{figure}
\resizebox{\hsize}{!}{\includegraphics[]{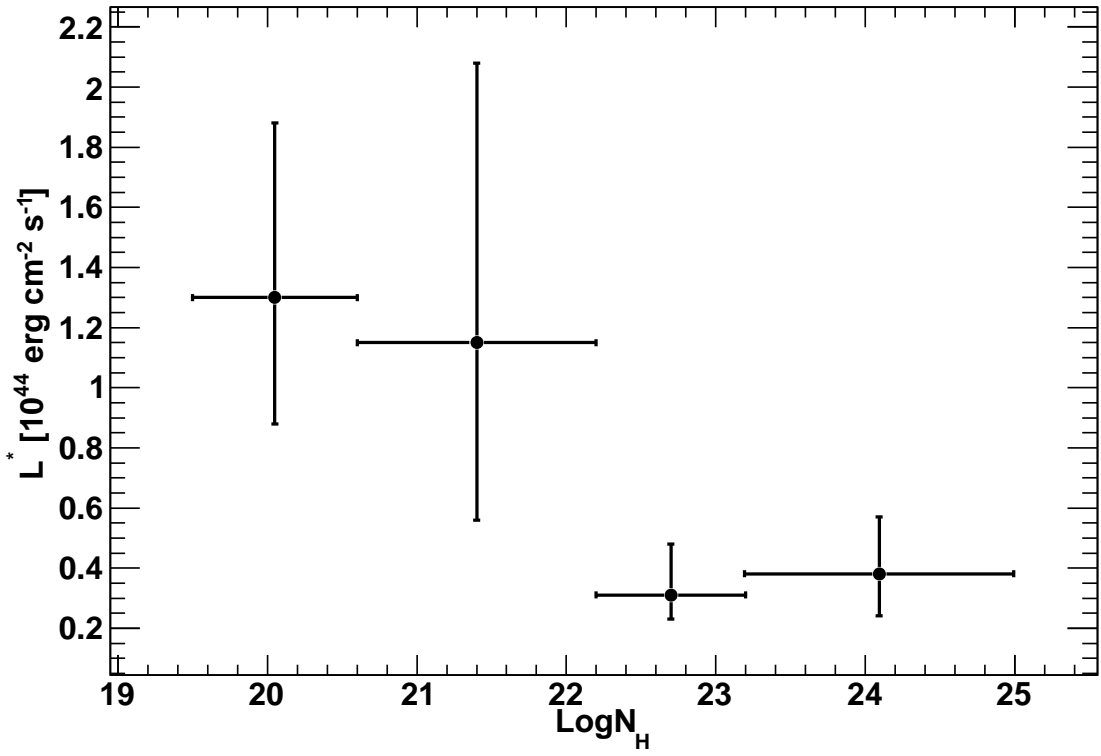} }
\caption{Break luminosity ($L^*$, as in Eq.~\ref{eq:2pow}) in units of 10$^{44}$ erg cm$^2$ s$^{-1}$, versus column density. The sample was divided into four equally populated subsamples of $\sim$50 sources. Errors are quoted at 1\,$\sigma$ level.
\label{fig:lbreak} }
\end{figure}

We tested if this 'trend' of the break luminosity holds for smaller subsamples, namely if the very most
absorbed sources show a significantly smaller value of $L^*$. This exercise 
needs of course a sufficient number of sources per subsample, in order to constrain the parameters of the fit with the 
broken power law.
Therefore we divided the parent population into four bins of absorption containing  $\sim50$ sources each. 
For each of the subsamples we computed the best fit values of the XLF as described before.
In Fig.~\ref{fig:lbreak} we showed the break luminosity (in units of 10$^{44}$ erg cm$^2$ s$^{-1}$) versus \nh. 
This exercise confirms, albeit with a statistical significant uncertainty, that absorbed AGN appear to be intrinsically less luminous
than unabsorbed AGN. 
 We also performed an additional test 
in order to exclude that this finding is partially driven
by the bias against the detection of the most absorbed AGN.
We computed the {\it intrinsic} XLF of absorbed AGN by de-absorbing
the AGN luminosities using the model described in $\S$\ref{sec:ct}.
As clearly seen from Fig.~\ref{fig:nh_intr2}, this correction is negligible
for all AGN with Log\nh$\leq$23.5, modest for all those with $23.5<$Log\nh$\leq$24
and relevant for AGN with Log\nh$\geq$24. It has to be noted that the
 {\it intrinsic} XLF of absorbed AGN suffers from incompleteness at the lowest
 luminosities. Indeed, because of the effect of large absorption, sources with
an intrinsic luminosity large enough to be detected by BAT might be pushed below
the BAT sensitivity. In order to avoid this problem we cut the sample
at the minimum de-absorbed luminosity for which the $K(N_H)$ correction
(discussed in $\S$\ref{sec:ct}) was less than 1. This minimum, de-absorbed,
luminosity is 2$\times10^{42}$\,erg s$^{-1}$ and we consider the BAT sample to be
complete above it. The parameters of the {\it intrinsic} XLF  of absorbed AGN
are reported in Tab.~\ref{tab:xlf_fits}. It is clear that the {\it intrinsic} XLF
of absorbed AGN is found to be 
in very good agreement with the XLF of absorbed AGN. This is because the bias
against the detection of absorbed AGN is relevant only for Log\nh$\geq 24$
and above this threshold the BAT sample contains very few objects.

\begin{figure}
\resizebox{\hsize}{!}{\includegraphics[]{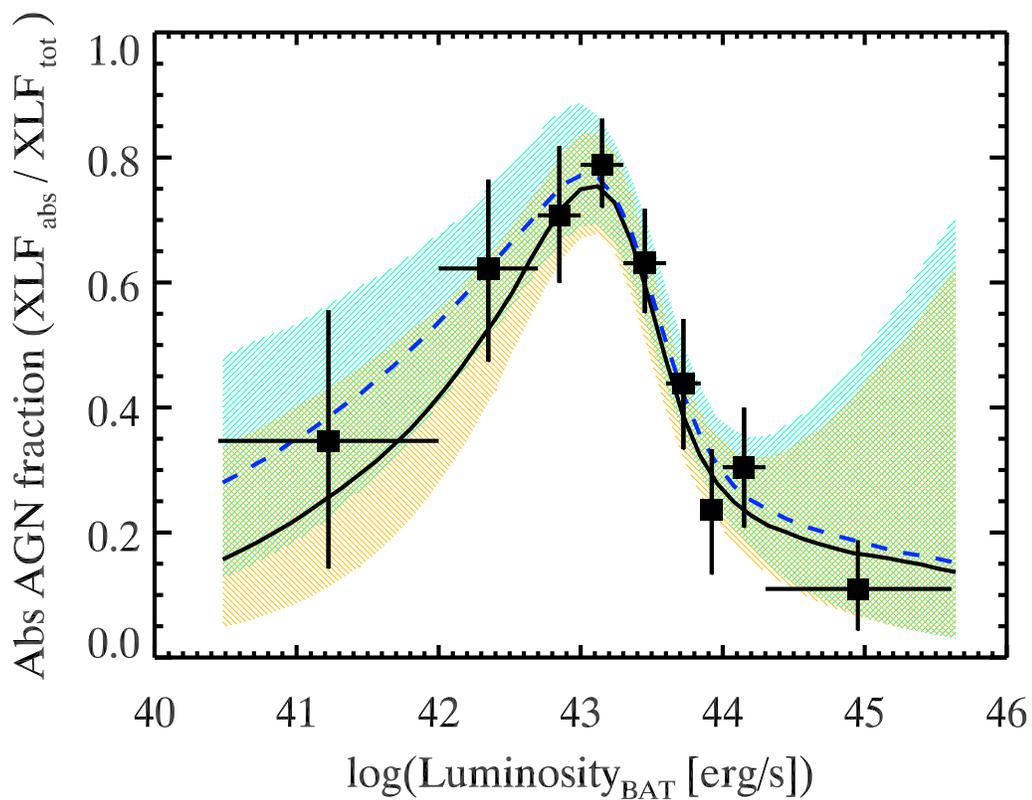} }
\caption{Fraction of the absorbed AGN versus luminosity in the 15--55 keV energy range. The lines are obtained by dividing the XLF of obscured AGN by the XLF  of the whole population.  The dashed (solid) line represents the whole sample (with a cut at $2\times 10^{42}$ erg s$^{-1}$).
The shaded bands represent the 1\,$\sigma$ uncertainty computed with a bootstrap analysis,  blue for the whole sample and yellow for the sample with the cut in luminosity.
\label{fig:xlf_boot_1000} }
\end{figure}

Finally, we test whether the anti-correlation of the fraction of obscured
AGN and luminosity (found in the previous section)  can be reproduced by the
two different XLFs for absorbed and unabsorbed AGN.
To this extent we divide the XLF of absorbed AGN by the XLF of the
entire sample. We computed the 1\,$\sigma$ error on this function 
via bootstrap with replacement
employing 1000 bootstrapped samples.  In order not to suffer from biases
derived from the detection of Compton-thick AGN and also to avoid the need
of correcting for them, we also excluded all objects with Log$N_{H}\geq$24.
As shown in the previous sections the BAT sample can be considered complete
for all AGN with Log$N_{H}\leq$24.
The results of our analysis are shown in Fig.~\ref{fig:xlf_boot_1000}
along with the observed anti-correlation of the obscured AGN fraction and
luminosity. 
 The agreement within errors is clear at all luminosities and
the decreasing trend in the absorbed fraction at low luminosity does not depend
on the presence or the absence of Compton-thick AGN.
The trend of the fraction of obscured AGN and luminosity 
 can be perfectly reproduced by the analysis of the XLFs
of the absorbed and unabsorbed AGN.
In turn, this difference can be
ascribed to the fact that on average obscured AGN appear to be less luminous.
Even more interestingly, the ratio of XLFs predicts (in agreement with the
observations) that the fraction of obscured sources decreases at low luminosities
(i.e. $L_X < 10^{42}$\,erg s$^{-1}$). However, we caution the reader that, due to the lack
of sources, the obscured AGN fraction is, at low-luminosities, compatible within 3\,$\sigma$
with a constant fraction of $\sim80$\,\%.

\section{Discussion}

\subsection{Compton-thick AGN and the Cosmic X-ray Background}
It is well understood that the shape and the intensity of the CXB cannot
be reproduced successfully if Compton-thick AGN (logN$_{\rm H}\geq 24$)
are not invoked. They are required to produce $\sim$10\,\%
of the CXB intensity at $\sim30$ keV \citep{gilli07,treister09}.
However, since the extreme absorption makes these objects
faint at X-rays,
Compton-thick AGN have to be numerous (e.g. $\sim$30\,\% of the whole AGN
population) in order to fulfill the above requirement.
Despite this general belief, all searches of Compton-thick AGN
have so far highlighted  a lack of Compton-thick AGN 
(see Tab.~\ref{tab:ctstory}). This work, which uses the largest AGN
sample  collected so far above 15 keV, shows that the {\it detected}
fraction of Compton-thick AGN is only 4.6$^{+2.1}_{-1.5}$\,\%.
At the fluxes currently sampled by \sw--BAT 
population synthesis models predict this fraction to be
either $\sim$16\% if one adopts the \cite{gilli07} or $\sim$8\%
the \cite{treister09} model. Thus our measurement appears to be substantially
lower than both predictions, but compatible within $\sim$2\,$\sigma$ with
what reported by \cite{treister09}.

However, one must take into account that even X-rays with energies larger
than 15\,keV are absorbed if the source is mildly Compton-thick.
Fig.~\ref{fig:nh_intr} shows how dramatic this effect is.
Indeed, if the source has logN$_{\rm H}\approx$24
then only $\sim$50\,\% of the intrinsic, nuclear, flux is visible above
15\,keV. This fraction becomes much lower (e.g. a few \%) if the 
source has logN$_{\rm H}\approx$25. Thus the extreme low number of Compton-thick
AGN detected in shallow surveys at hard X-rays is due to fact that 
only the population of {\it intrinsically} bright 
objects is being sampled. 
As such these objects are rare. For the first time, we use
the knowledge of how X-rays are absorbed through a Compton-thick medium
to determine the {\it intrinsic} absorption distribution. We find
that Compton-thick AGN are 20$^{+9}_{-6}$\,\% (statistical
error) of the whole AGN population. Our result shows beyond doubt
that Compton-thick sources are indeed {\it intrinsically} numerous, but
due to the large absorption, very difficult to detect. 

The average spectrum of all Compton-thick AGN detected by BAT is, in
the 15-195\,keV band, compatible with a very curved spectrum peaking
around 40--50\,keV (see Fig.~\ref{fig:agn_stack}).
Since the peak of the CXB is located at $\sim25$\,keV \citep{ajello08c},
this implies that the bulk of the Compton-thick AGN must be located
at z$\approx 1$ in order to explain the missing fraction of the CXB.
 We employed the MYTorus model,
 that fully treats photoelectric
absorption and relativistic Compton-scattering in a toroidal
geometry,  to fit the BAT spectra of the Compton-thick AGN.
We derived that in the BAT band the spectrum of Compton-thick
AGN is very likely dominated by the scattered component. The photon index
is compatible with 2.0 while the absorbing
column density is constrained to be Compton-thick using BAT data alone.
We also found out that on average only $\sim$30\,\% of the intrinsic
nuclear flux is visible in the BAT band. A more detailed an analysis
on each of the Compton-thick AGN detected by BAT will be reported in
a future publication.


Summarizing we find that Compton-thick AGN: 1) are {\it intrinsically}
as numerous as required by the AGN unified model ,
and 2) have a spectral shape which 'fits' the one required 
to explain the shape and the peak of the CXB. However, the predictions
 of population synthesis models \citep{gilli07,treister09}
in terms of {\it observed} Compton-thick AGN are between a factor
of 2 to 3 larger than what is currently observed by BAT.


\subsection{A receding torus or a Clumpy torus model ?}

It is generally accepted that all kinds of AGN are different 
manifestations of the same engine (i.e. the accreting
super-massive black hole), seen under different line of sights.
Thus, in the unified picture, 
obscuration depends solely on orientation effects.
This hypothesis breaks down when showing that the fraction
of obscured sources (relative to the whole population) decreases
with luminosity. This result has been reported
several times in the past, but it has been confirmed now by BAT (with a sample a factor $\sim$2 greater than the one presented in \cite{tueller08}).

 Our results, like the previous claims of this anti-correlation, 
are inconsistent with the simplest prediction of the unified model, 
for which $F_{\%}$ should not depend on L$_X$. 
Already \cite{lawrence91} tried to overcome this contradiction 
by proposing a \emph{receding torus model}.  In this model, 
the inner radius is set by the dust sublimation radius which increases
with source luminosity. This model predicts that the fraction of obscured
sources scales with luminosity as $F_{\%}\propto L^{-0.5}$.

In basic terms \cite[see][for a detailed discussion]{nenkova08}, the inner radius is set from the following:
\begin{equation}
R_d \simeq 0.4 \left( \frac{L}{ 10^{45} \rm{erg\,\ s^{-1}} }\right)^{0.5} \left(\frac{1500~\rm{K}}{T_{sub}}\right)^{2.6} \rm{pc}
\end{equation}
where L is the bolometric luminosity and $T_{sub}$ is the sublimation temperature of the dust. Given the angular dependence of $F_{\%}$ (i.e. the angle
at which a smooth torus becomes transparent to radiation), a constant H/R implies $F_{\%} \propto L^{-0.5}$.
While attractive, this simple idea fails to reproduce correctly the dependence
of the obscured fraction with luminosity, in particular when looking just at the hard X-ray portion of the spectral energy distribution. 
Indeed, in this work we showed that $F_{\%}\propto L_X^{-0.26\pm0.05}$ for L$_X\geq$10$^{42}$\,erg s$^{-1}$.

Nonetheless, one has to consider that (i) the torus is likely clumpy, (ii)  H/R is a function of luminosity \citep{simpson05}, and 
(iii) that even if the dusty components of the torus do absorb optical, UV, and X-ray
photons, X-ray obscuration  can take place  also in the -dust free- inner region, in the proximity ($<0.1$pc) of the AGN. This has been largely discussed in 
\cite{risaliti02,risaliti07}.
Recently \citep[see][]{hoenig07,nenkova08} $F_{\%}$ was interpreted in the 
framework of a \emph{radiation-limited clumpy dust torus} in which $F_{\%}\propto L^{-0.25}$, which is much closer to our findings.
In this scenario, the obscuration--luminosity relation is recovered in terms of probability that the photons intercept a sufficient number of clouds in the line 
of sight. 
In addition, the work of \cite{hasinger08} shows similar results in the 2--10 keV energy band, even if just proxies of \nh\ are used. 
Finally, \cite{maiolino07} interpreted those early results by comparing them to a similar relation 
between the fraction of obscured AGN and the O[III] luminosity (as well as other bands, but see their detailed description).
According to the authors this was a signature of the variation of the covering factor of the AGN dust with luminosity.
 The variation of the covering factor of the torus is -in general- also associated with the ``Iwasawa-Taniguchi'' (IT) effect \citep{iwasawa93}, i.e. the anti-correlation of the EW of FeK$\alpha$ and X-ray luminosity. \cite{bianchi09}  discussed the agreement among the findings of \cite{dadina08} in the 20-100 keV regime, and their updated \xmm\ sample. If the anti-correlation of $F_{\%}$ and luminosity is interpreted as the variation of the covering factor of the torus, then this can be in turn compared to the IT effect. From the results of \cite{dellaceca08}, they found a slope of $\simeq 0.22$, in very good agreement with the IT effect, with the slope ($\simeq 0.18$) presented in \cite{maiolino07}, and eventually with our findings ($\simeq 0.26$). Although these similarities are intriguing, they certainly deserve a profound investigation (which is beyond the goals of this paper). This is even more compelling when considering that the methods differ and in principle sample different materials. 


In the low luminosity regime (i.e. below 10$^{42}$\,erg s$^{-1}$), 
the behavior of the torus is likely more complex than what reported above.
Different authors \cite[e.g.][]{elitzur06,elitzur09} showed that below 
a bolometric luminosity of 
$\sim 10^{42}$\,erg s$^{-1}$ the torus obscuration region disappears.
This results under the assumption that the clouds are generated by a disk-wind outflow, rather then accreted from the galaxy 
\citep{krolik88}.
In the framework proposed by \cite{elitzur06} and \cite{elitzur09} the cloud mass outflow $\dot{M}_{cl}$ is proportional to the mass accretion rate, 
which is related to the bolometric luminosity
via $\dot{M}_{acc} = 0.02 L_{45} / \eta M_{\sun}$ yr$^{-1}$, $\eta$ being the accretion efficiency. 
The ratio $\dot{M}_{cl}/\dot{M}_{acc}$ increases at progressively lower luminosities but since it 
cannot exceed unity, 
there is a limiting luminosity at which the system cannot sustain the cloud outflow any longer. 

At low luminosities we then expect absorption of X-rays to be less effective, due to the lack of obscuration in the torus. We should then
observe a decrease in the fraction of obscured AGN at low luminosity.
For the first time, and thanks to BAT,
we are able to inspect the behavior of $F_{\%}$ at low
luminosities. Indeed, Fig.~\ref{fig:xlf_boot_1000} shows\footnote{We caution the reader that within 3$\sigma$ the behavior
of $F_{\%}$ at low luminosity is still compatible with a constant value of 0.8.} 
that the fraction
of obscured AGN decreases at low luminosities as one would expect if the
torus obscuring region would cease to exist.
There are no other evidences (perhaps beside this one, but see \citealp{vanderwolk10}) which shows that the torus
disappears at low luminosities, but there are ample evidences that 
at least the broad line region (BLR) disappears at low luminosities.
This happens for two known class of objects: 
1) BL Lac sources and 2) low-luminosity type-2 AGN.

BL Lacs are a class of (low-luminosity)
blazars characterized by the absence of emission lines
in their optical spectrum. Their broad-band spectrum is normally well
understood in terms of the Synchrotron-Self Compton model \citep{maraschi92} where
the electrons responsible for the synchrotron emission are up-scattering
(via inverse Compton) the same synchrotron radiation to high energy. 
The main difference with the more luminous flat spectrum radio quasars
is that in these latter ones an additional high-energy component
(refereed to as 'external Compton' component) is normally detected. 
In the external Compton model photons from the BLR and/or the disk are up-scattered
to high energy by the electrons in the jet. It is believed that
 the absence of lines and of this external components is caused, 
in BL Lac objects, by the lack of the BLR, and a lower radiation field density.
Moreover, BL Lacs are characterized by low Eddington ratios
(e.g. $\lambda_{Edd}\simeq 0.01$).
Around this Eddington ratio, the accretion process experiences a transition
from an optically thick, geometrically thin disk ($\lambda_{Edd}>0.01$) 
to a radiatively inefficient geometrically thick disk \cite[][]{ghisellini09}.
Therefore in blazars with low Eddington ratios,
 the electrons in the jet find a medium starved of external radiation, 
weak or no lines are produced, and the AGN is classified as BL Lac.

Another evidence for the absence of BLR at low luminosities is produced
by low-luminosity type-2 AGN. These AGN, are sometimes referred to
as 'true' type-2 AGN because when observed in polarized light they
do not show broad lines \citep{ho08} and hence lack a BLR.
How the ``local'' changes in the 
accretion regime affect the environment at the BLR and the torus region is non-trivial and debated  (according to \cite{nicastro00} and \cite{nicastro03}, the main driver for the disappearance of the BLR appears to be the accretion rate), nonetheless in this framework 
we find a convincing interpretation of our findings.
The clumpy torus, and the BLR progressively disappear, resulting in a less efficient X-ray obscuration. 
Therefore we should expect $F_{\%}$ to flatten or even to invert its dependence on the luminosity at progressively low Eddington ratios.
With future X-ray missions like {\it NuSTAR} \citep{harrison05},
and {\it NHXM} \citep{pareschi09} it will be possible to sample
with better statistics the population of the low-luminosity AGN
and to investigate the behavior of the BLR and the torus in greater detail.

\section{Summary and conclusion}
We addressed the study of a complete, flux limited, sample of local AGN collected by the \sw--BAT instrument in the first three
years of survey. The sources are listed in Tab.\,\ref{tabellon}, along with their properties. 
The aim of this work is to characterize the AGN population from two fundamental observables such as the hard X-ray
(15--55 keV) luminosity and absorption. To this aim we jointly fitted the BAT spectra with the available follow up in the 0.3--10 keV domain. 
In the following we briefly review the main findings. 
We remind the reader that AGN are defined 'absorbed' if the column density for photoelectric absorption exceeds 10$^{22}$ atoms cm$^{-2}$.
\begin{itemize}

\item	Performing a stacked analysis of the complete 199 AGN sample, a simple power law model was shown not to account for the continuum 
		emission. In addition we performed the stacked analysis of the different subsamples of sources: unabsorbed, absorbed, and 
		-for the first time- Compton-thick one. 
 The average spectrum of CT sources was found to be dominated (in the BAT band) by the scattered component and its photon index was found
to be compatible with $\sim$2.0. According to our results, only $\sim$30\,\%
of the source intrinsic flux is visible in the 15--55\,keV band.		
		
\item	We showed that absorbed AGN are characterized by slightly harder spectra (1.91) with respect to the unabsorbed ones (2.00). 
		Nonetheless the distributions are quite broad, resulting in a Kolmogorov-Smirnov probability of $3.5\times 10^{-3}$ 
		of belonging to the same parent population.
		
\item	We computed the observed \nh\ distribution, which shows that the observed fraction absorbed sources is 54$\pm4$\%, with columns peaking 
		at 10$^{23}$ cm$^{-2}$. The observed fraction of Compton-thick objects is $4.6 ^{+2.1}_{-1.5} \%$, a factor 2 to 3 lower than what predicted from 
		population synthesis models at the fluxes of the BAT survey.
		
\item	We estimated the bias of the BAT instrument against the detection of the Compton-thick AGN.
		We consequently derived the \emph{intrinsic} \nh\ distribution by integrating the logN-logS in bins of logN$_{\rm H}$
		setting the minimum observed flux of integration so that the limiting intrinsic flux was the same for all the bins.
		Therefore we showed that even if the CT objects are only a minor fraction 
		of the observed sample, their contribution rises to 20$^{+9}_{-6}$\,\% in the intrinsic AGN population.
		
\item 	The relation between the observed fraction of obscured AGN and the hard X-ray luminosity ($F_{\%}$),
		was found to have different behaviors according to the luminosity regime considered.
		For luminosities greater then 10$^{42}$ erg s$^{-1}$ we found a monotonic
		decline with a slope of -0.26$\pm 0.05$. At smaller luminosities, albeit affected by poor statistics,
		we found a flattening of $F_{\%}$ which we interpreted as the manifest disappearance of the obscuring region. In a disk--cloud 
		outflow scenario, this is indeed expected to happen under a critical luminosity, which is of the same order of the luminosity 
		at which we observe the flattening.
		 
\item	We showed that the obscuration--luminosity relation can be explained by the different X-ray Luminosity Functions of the obscured and unobscured
		subsamples. This in turn means that absorbed AGN are intrinsically less luminous.
		This result, if the mass distribution is narrow, points towards a trend in the Eddington ratios in which objects accreting 
		at lower values (thus having smaller effects on their environments) are more absorbed. Obscuring clouds would be able to come 
		closer to the nuclear region without being affected, and bury the AGN. 

\end{itemize}

A key test to improve our findings will be the calculation of the BH masses for the sample. By means of the two physical quantities (i.e. 
$\lambda_{Edd}$ and mass) we plan to test whether it is possible to find a sequence relating absorption, Eddington ratio, and mass of the 
black hole. Some information can already be found in \cite{middleton08} and \cite{beckmann09}. They found that for two different
samples of AGN selected in the local Universe, the mean Eddington ratio
is in the 0.01-0.06 range. \cite{cap10} also
showed that, on average, the BAT AGN used in this work have an Eddington
ratio of 0.01. Moreover, \cite{middleton08} and \cite{beckmann09} found
that on average
unobscured AGN have larger Eddington ratios with respect to obscured ones. This would be consistent with the presence of a trend in $\lambda_{Edd}$ for non-jetted AGN.
The obscured objects could be accreting at lower Eddington ratios and with flatter spectra; the unobscured ones, with steeper 
spectra, could be accreting slightly more efficiently. 
By means of this analysis it would be also possible to relate our findings to a physical consistent picture, in the framework of 
merger driven AGN activity. Indeed, merging of gas-rich
galaxies provides an efficient way to funnel large amount of gas and dust
to the central black hole  and 
triggers AGN activity \cite[e.g.][]{kauffmann00,
wyithe03,croton06}. Noteworthy a very recent work by \cite{koss10} determined that a considerable fraction 
(i.e. $\sim$24\,\%)  of the hosts of the BAT AGN have a close companion
within 30\,kpc and are experiencing a major merging event. This fraction is extremely relevant when compared to 
a control sample of local (i.e. $z<0.1$) optically selected narrow-line AGN, where the fraction of interacting companions is $\sim1\,\%$. 
We counted how many objects of the sample of merging AGN of \cite{koss10}
are obscured. We found that $\geq$63\,\% are obscured AGN\footnote{
For comparison, we remind that in our sample the fraction of obscured
sources is $\sim$50\,\%.}
with an average column density of logN$_{\rm H}\approx23.4$. 
A limitation to this simple picture is that in case of a merging
event and of a large gas quantity being funneled towards the center, the
black hole is expected to accrete with  high Eddington ratios
\citep[e.g.][]{dimatteo05,hopkins06}.
According to \cite{fabian99} a SMBH accreting at Eddington luminosities
should clear the environment from any Compton-thin (e.g. logN$_{H}<24$)
column density and therefore transit to a less obscured phase.
If the very first phase of the gas-rich merger event is the creation
of a Compton-thick AGN, then we would expect it to display large
Eddington ratios. However, two of the most famous Compton-thick AGN
(Circinus and NGC 4945) display an Eddington ratio far from unity 
\citep[i.e. $\leq10^{-2}$, see][]{gultekin09}, proving that this argument still escapes a conclusive explanation.
The largest hard X-ray selected samples of AGN in the local Universe may shed some light 
on the physical interpretation of the feedback of black holes on their surroundings.


\begin{acknowledgements}
We are in debt to Tahir Yaqoob and Kendrah Murphy
for allowing us to use the results of their model for the  transmission of radiation
through a Compton-thick medium before publication. We also acknoweldge helpful comments from the referee.
The authors acknowledge the use of NED, SIMBAD, and  HEASARC. We thank the \sw\ team for the rapid approval of ToO observations.
D.B. is in debt to  G. Ghisellini and G. Ghirlanda for endless discussions, 
acknowledges S. Sazonov for stimulating discussions on the dusty torus and the luminosity function,
and M. Bolzonella for help with the Binomial statistics computation of errors. D.B. also acknowledges R. Gilli, E. Treister, and P. Severgnini for their kind replies.
D.B. is supported through DLR 50 OR 0405. A.C. acknowledges support from the following: ASI-INAF I/23/05 and ASI-INAF I/088/06/0.
\end{acknowledgements}

{\it Facilities:} \facility{Swift/BAT}, \facility{Swift/XRT}, \facility{XMM/Newton}.



\clearpage
\begin{deluxetable}{lccccllccll}
\tablewidth{0pt}
\tabletypesize{\scriptsize}
\rotate
\tablecaption{\sw\ sample of AGN. \label{tabellon}}
\tablehead{
\colhead{SWIFT NAME}               & \colhead{R.A.}       &
\colhead{DEC}            & \colhead{Flux [15-55 keV]}     &
\colhead{S/N}            & \colhead{ID}    & \colhead{Type$^{\ddag}$} & 
\colhead{redshift}         & \colhead{Ph. Index}     & 
\colhead{log(N$_{{\rm H}}$) }     & \colhead{Reference} \\
\colhead{}               & \colhead{\small (J2000)}  &
\colhead{\small (J2000)}       & \colhead{\small(10$^{-11}$ cgs)} &
\colhead{}               & \colhead{}         & \colhead{} & \colhead{} &
\colhead{(BAT only)}
}

\startdata

J0006.4+2009 & 1.600 & 20.152 & 1.16$\pm0.20$ & 5.8 & Mrk 335 
& Sy1 & 0.03 &  2.38$^{+0.50}_{-0.42}$ & 22.6&(2)\\ 
J0038.6+2336 & 9.650 & 23.600 & 1.10$\pm0.21$ & 5.3 & Mrk 344 
& Sy & 0.02 &  2.06$^{+0.46}_{-0.41}$ & 23.2&(2)\\ 
J0042.7-2332 & 10.680 & -23.548 & 2.44$\pm0.21$ & 11.7 & NGC 235A
 & Sy2 & 0.02 &  1.60$^{+0.20}_{-0.19}$ & 23.0&(1)\\ 
J0048.7+3157 & 12.188 & 31.962 & 7.71$\pm0.20$ & 37.8 & Mrk 348
 & Sy2 & 0.02 &  1.90$^{+0.08}_{-0.07}$ & 23.3&(1)\\ 
J0051.9+1726 & 12.998 & 17.447 & 1.81$\pm0.21$ & 8.6 & QSO B0049+171
 & Sy1 & 0.06 &  2.13$^{+0.28}_{-0.25}$ & 20.0&(2)\\ 
J0059.9+3149 & 14.997 & 31.831 & 1.66$\pm0.21$ & 8.0 & SWIFT J0059.4+3150
 & Sy1.2 & 0.01 &  1.93$^{+0.35}_{-0.32}$ & 21.0&(1)\\ 
J0101.0-4748 & 15.274 & -47.800 & 0.97$\pm0.18$ & 5.6 & 2MASX J01003469-4748303
 & GALAXY & 0.08 &  2.32$^{+0.44}_{-0.38}$ & 22.6&(2)\\ 
J0108.8+1321 & 17.201 & 13.351 & 1.78$\pm0.22$ & 8.2 & 4C 13.07
 & Sy2 & 0.06 &  1.76$^{+0.34}_{-0.27}$ & 23.8&(3)\\ 
J0111.4-3805 & 17.867 & -38.086 & 1.52$\pm0.18$ & 8.3 & \bf{NGC 424}
 & Sy2 & 0.01 &  1.94$^{+0.28}_{-0.27}$ & 24.3 &(4)\\ 
J0113.8-1450 & 18.453 & -14.850 & 1.24$\pm0.21$ & 5.8 & Mrk 1152
 & Sy1 & 0.05 &  2.10$^{+0.38}_{-0.34}$ & 21.1&(5)\\ 
J0114.3-5524 & 18.600 & -55.400 & 0.92$\pm0.17$ & 5.3 & SWIFT J0114.4-5522
 & Sy2 & 0.01 &  1.53$^{+0.33}_{-0.38}$ & 22.9&(1)\\ 
J0123.8-5847 & 20.952 & -58.785 & 2.65$\pm0.17$ & 15.3 & Fairall 9
 & Sy1 & 0.05 &  2.02$^{+0.15}_{-0.15}$ & 20.4&(1)\\ 
J0123.8-3504 & 20.974 & -35.067 & 2.72$\pm0.18$ & 14.7 & NGC 526A
 & Sy1.5 & 0.02 &  1.71$^{+0.14}_{-0.14}$ & 22.3&(1)\\ 
J0127.9-1850 & 22.000 & -18.847 & 1.27$\pm0.20$ & 6.2 & MCG-03-04-072
 & Sy1 & 0.04 &  2.26$^{+0.54}_{-0.44}$ & 20.0&(2)\\ 
J0134.0-3629 & 23.506 & -36.486 & 2.36$\pm0.18$ & 13.0 & NGC 612
 & Sy2 & 0.03 &  1.63$^{+0.17}_{-0.15}$ & 23.7&(6)\\ 
J0138.6-4000 & 24.674 & -40.008 & 3.17$\pm0.18$ & 18.0 & ESO 297-018
 & Sy2 & 0.03 &  1.71$^{+0.11}_{-0.11}$ & 23.8&(7)\\ 
J0142.6+0118 & 25.652 & 1.300 & 1.28$\pm0.22$ & 5.7 & [VV2003c] J014214.0+011615
 & Sy1 & 0.05 &  2.52$^{+1.10}_{-0.57}$ & \nodata\\ 
J0152.9-0326 & 28.250 & -3.448 & 1.47$\pm0.22$ & 6.6 & IGR J01528-0326
 & Sy2 & 0.02 &  2.28$^{+0.48}_{-0.41}$ & 22.9&(2)\\ 
J0201.2-0649 & 30.320 & -6.821 & 4.17$\pm0.22$ & 19.3 & NGC 788
 & Sy2 & 0.01 &  1.74$^{+0.11}_{-0.11}$ & 23.5&(1)\\ 
J0206.5-0016 & 31.631 & -0.270 & 1.53$\pm0.22$ & 6.9 & MRK 1018
 & Sy1.5 & 0.04 &  1.48$^{+0.40}_{-0.39}$ & 20.5&(1)\\ 
J0215.0-0044 & 33.751 & -0.749 & 1.30$\pm0.22$ & 5.9 & Mrk 590
 & Sy1.2 & 0.03 &  2.23$^{+0.54}_{-0.47}$ & 20.4&(1)\\ 
J0226.0-6315 & 36.500 & -63.250 & 0.91$\pm0.18$ & 5.2 & FAIRALL 0926
 & Sy1 & 0.06 &  2.55$^{+0.57}_{-0.47}$ & 20.9 &(2)\\ 
J0226.8-2819 & 36.703 & -28.324 & 1.14$\pm0.18$ & 6.4 & 2MASX J02262568-2820588
 & Sy1 & 0.06 &  2.21$^{+0.59}_{-0.49}$ & 21.8&(2)\\ 
J0228.4+3118 & 37.120 & 31.316 & 4.38$\pm0.23$ & 19.4 & NGC 931
 & Sy1.5 & 0.02 &  2.25$^{+0.22}_{-0.16}$ & 21.6&(1)\\ 
J0232.0-3639 & 38.020 & -36.662 & 1.09$\pm0.17$ & 6.4 & IC 1816
 & Sy2 & 0.02 &  2.03$^{+0.38}_{-0.34}$ & 23.9&(2)\\ 
J0234.4+3229 & 38.612 & 32.489 & 1.60$\pm0.23$ & 7.1 & NGC 973
 & Sy2 & 0.02 &  1.70$^{+0.41}_{-0.33}$ & 22.5&(2)\\ 
J0234.8-0847 & 38.702 & -8.794 & 2.13$\pm0.21$ & 10.2 & NGC 985
 & Sy1 & 0.04 &  2.23$^{+0.26}_{-0.24}$ & 21.6&(1)\\ 
J0235.6-2935 & 38.900 & -29.600 & 0.99$\pm0.18$ & 5.6 & ESO 0416-G0002
 & Sy1.9 & 0.06 &  1.62$^{+0.45}_{-0.35}$ & 19.6&(1)\\ 
J0238.5-5213 & 39.647 & -52.220 & 1.31$\pm0.17$ & 7.6 & ESO 198-024
 & Sy1 & 0.05 &  1.69$^{+0.26}_{-0.25}$ & 21.0&(1)\\ 
J0239.0-4043 & 39.767 & -40.732 & 0.97$\pm0.17$ & 5.8 & 2MASX J02384897-4038377
  & Sy1 & 0.06 &  2.12$^{+0.61}_{-0.52}$ & 20.0&(2)\\ 
J0241.5-0813 & 40.381 & -8.220 & 1.34$\pm0.21$ & 6.4 & NGC 1052
 & Sy2 & 0.01 &  1.47$^{+0.38}_{-0.38}$ & 20.5&(8)\\ 
J0242.9-0000 & 40.732 & -0.012 & 2.00$\pm0.22$ & 8.9 & \bf{NGC 1068}
 & Sy2 & 0.004 &  2.23$^{+0.33}_{-0.30}$ & $>$25&(4)\\ 
J0249.3+2627 & 42.349 & 26.451 & 1.25$\pm0.23$ & 5.5 & IRAS 02461+2618 & Sy2
 & 0.06 &  1.66$^{+0.39}_{-0.38}$ & 23.5&(2)\\ 
J0252.8-0830 & 43.200 & -8.500 & 1.06$\pm0.21$ & 5.0 & MCG-02-08-014
 & Sy2 & 0.02 &  1.69$^{+0.54}_{-0.40}$ & 23.1&(2)\\ 
J0255.4-0010 & 43.873 & -0.170 & 4.48$\pm0.22$ & 20.1 & NGC 1142
 & Sy2 & 0.03 &  1.85$^{+0.12}_{-0.11}$ & 23.4&(9)\\ 
J0256.4-3212 & 44.117 & -32.208 & 1.31$\pm0.17$ & 7.7 & ESO 417-6
 & Sy2 & 0.02 &  1.86$^{+0.25}_{-0.24}$ & 22.9&(2)\\ 
J0311.6-2045 & 47.919 & -20.760 & 1.27$\pm0.18$ & 6.9 & 2MASX J03111883-2046184
 & Sy1 & 0.07 &  1.79$^{+0.43}_{-0.34}$ & 20.0&(2)\\ 
J0325.1+3409 & 51.296 & 34.152 & 1.61$\pm0.24$ & 6.8 & 2MASX J03244119+3410459
 & Sy1 & 0.06 &  1.56$^{+0.42}_{-0.30}$ & 20.0&(2)\\ 
J0333.5+3716 & 53.397 & 37.278 & 1.63$\pm0.24$ & 6.8 & IGR J03334+3718
 & Sy1 & 0.06 &  2.37$^{+0.53}_{-0.44}$ & 20.0&(2)\\ 
J0333.7-3608 & 53.433 & -36.141 & 3.12$\pm0.17$ & 18.9 & NGC 1365
 & Sy1.8 & 0.01 &  2.02$^{+0.25}_{-0.24}$ & 23.6&(1)\\ 
J0342.2-2114 & 55.554 & -21.244 & 2.15$\pm0.18$ & 11.8 & SWIFT J0342.0-2115
 & Sy1 & 0.01 &  1.88$^{+0.20}_{-0.19}$ & 20.5&(1)\\ 
J0347.3-3029 & 56.850 & -30.500 & 0.89$\pm0.17$ & 5.3 & RBS 0741
 & Sy1 & 0.10 &  2.04$^{+0.48}_{-0.43}$ & 20.4 & (2)\\ 
J0350.7-5022 & 57.679 & -50.377 & 1.29$\pm0.17$ & 7.5 & SWIFT J0350.1-5019
 & Sy2 & 0.04 &  1.79$^{+0.39}_{-0.29}$ & 23.2&(2)\\ 
J0357.0-4039 & 59.268 & -40.666 & 0.89$\pm0.17$ & 5.4 & 2MASX J03565655-4041453
 & Sy1.9 & 0.07 &  1.56$^{+0.35}_{-0.34}$ & 22.5&(1)\\ 
J0402.5-1804 & 60.639 & -18.077 & 1.35$\pm0.19$ & 7.0 & ESO 549- G049
 & Sy2 & 0.03 &  1.75$^{+0.50}_{-0.38}$ & 22.4&(2)\\ 
J0407.5+0342 & 61.883 & 3.717 & 1.90$\pm0.25$ & 7.6 & 3C 105
 & Sy2 & 0.09 &  1.91$^{+0.30}_{-0.28}$ & 23.4&(1)\\ 
J0415.2-0753 & 63.800 & -7.900 & 1.31$\pm0.23$ & 5.6 & LEDA 14727
 & Sy1 & 0.04 &  2.25$^{+0.44}_{-0.39}$ & 23.5&(2)\\ 
J0426.4-5712 & 66.603 & -57.201 & 1.40$\pm0.17$ & 8.2 & 1H 0419-577
 & Sy1 & 0.10 &  2.57$^{+0.56}_{-0.33}$ & 19.5&(1)\\ 
J0433.4+0521 & 68.355 & 5.365 & 5.21$\pm0.26$ & 19.8 & 3C-120
 & Sy1 & 0.03 &  2.12$^{+0.13}_{-0.12}$ & 21.2&(1)\\ 
J0438.5-1049 & 69.633 & -10.830 & 1.48$\pm0.23$ & 6.4 & MCG-02-12-050
& Sy1 & 0.04 &  2.02$^{+0.61}_{-0.53}$ & 20.0&(2)\\ 
J0444.7-2812 & 71.199 & -28.200 & 1.07$\pm0.18$ & 5.9 & 2MASX J04450628-2820284
 & Sy2 & 0.15 &  2.17$^{+0.88}_{-0.66}$ & 20.0&(2)\\ 
J0451.8-5807 & 72.966 & -58.133 & 0.88$\pm0.17$ & 5.2 & RBS 0594
 & Sy1 & 0.09 &  1.86$^{+0.42}_{-0.38}$ & 20.0&(2)\\ 
J0453.5+0403 & 73.380 & 4.060 & 2.11$\pm0.28$ & 7.6 & {\bf CGCG 420-015}
 & Sy2 & 0.03 &  2.04$^{+0.36}_{-0.33}$ & $\sim$24.2&(4)\\ 
J0455.3-7528 & 73.841 & -75.477 & 1.27$\pm0.18$ & 6.9 & ESO 33-2
 & Sy2 & 0.02 &  2.52$^{+0.49}_{-0.41}$ & 22.1 & (2)\\ 
J0505.9-2351 & 76.497 & -23.854 & 2.78$\pm0.20$ & 13.9 & XSS J05054-2348
 & Sy2 & 0.04 &  1.79$^{+0.18}_{-0.17}$ & 22.7&(1)\\ 
J0516.2-0009 & 79.071 & -0.161 & 4.11$\pm0.28$ & 14.5 & QSO B0513-002
 & Sy1 & 0.03 &  2.16$^{+0.17}_{-0.16}$ & 20.0&(2)\\ 
J0519.7-3240 & 79.930 & -32.676 & 2.38$\pm0.19$ & 12.9 & SWIFT J0519.5-3140
 & Sy2 & 0.04 &  1.72$^{+0.18}_{-0.14}$ & 21.1&(1)\\ 
J0519.8-4546 & 79.963 & -45.774 & 2.49$\pm0.17$ & 14.9 & Pictor-A
 & Sy1 & 0.04 &  1.90$^{+0.17}_{-0.17}$ & 21.0&(1)\\ 
J0524.2-1212 & 81.050 & -12.200 & 1.42$\pm0.25$ & 5.6 & LEDA 17233
 & Sy1 & 0.05 &  1.90$^{+0.36}_{-0.33}$ & 21.2&(2)\\ 
J0552.3-0727 & 88.090 & -7.457 & 14.75$\pm0.29$ & 51.7 & NGC 2110
 & Sy2 & 0.01 &  1.79$^{+0.00}_{-0.00}$ & 22.6&(1)\\ 
J0552.3+5929 & 88.100 & 59.500 & 1.14$\pm0.21$ & 5.3 & IRAS 05480+5927
 & Sy1 & 0.06 &  3.44$^{+1.00}_{-0.73}$ & 21.1&(2)\\ 
J0558.1-3820 & 89.549 & -38.347 & 2.12$\pm0.18$ & 11.6 & EXO 055620-3820.2
 & Sy1 & 0.03 &  2.21$^{+0.23}_{-0.21}$ & 22.2&(1)\\ 
J0602.9-8633 & 90.749 & -86.555 & 1.82$\pm0.22$ & 8.4 & \bf{SWIFT J0601.9-8636}
 & Sy2 & 0.01 &  1.67$^{+0.24}_{-0.37}$ & $\sim$24&(4)\\ 
J0603.1+6523 & 90.799 & 65.399 & 1.38$\pm0.20$ & 6.8 & UGC 3386
 & GALAXY & 0.02 &  2.10$^{+0.36}_{-0.33}$ & 23.2&(3)\\ 
J0615.8+7101 & 93.967 & 71.021 & 6.08$\pm0.20$ & 30.8 & Mrk 3
 & Sy2 & 0.01 &  1.66$^{+0.01}_{-0.01}$ & 24.0&(1)\\ 
J0623.9-3214 & 95.994 & -32.248 & 1.53$\pm0.20$ & 7.5 & ESO 426-G 002
 & Sy2 & 0.02 &  1.86$^{+0.38}_{-0.35}$ & 23.9&(2)\\ 
J0624.1-6059 & 96.028 & -60.998 & 1.25$\pm0.17$ & 7.4 & SWIFT J2141.0+1603
 & Sy2 & 0.04 &  2.51$^{+0.42}_{-0.36}$ & 23.4&(2)\\ 
J0640.7-4324 & 100.200 & -43.400 & 0.92$\pm0.18$ & 5.2 & 2MASX J06400609-4327591
 & Sy2 & 0.06 &  2.06$^{+0.44}_{-0.39}$ & 23.4&(2)\\ 
J0652.1+7425 & 103.044 & 74.425 & 3.29$\pm0.19$ & 17.1 & Mrk 6
 & Sy1.5 & 0.02 &  1.89$^{+0.13}_{-0.13}$ & 23.0&(1)\\ 
J0656.1+3959 & 104.027 & 39.986 & 2.29$\pm0.26$ & 8.7 & UGC 3601
 & Sy1 & 0.02 &  1.97$^{+0.29}_{-0.27}$ & 21.3&(2)\\ 
J0718.0+4405 & 109.517 & 44.084 & 1.67$\pm0.24$ & 7.1 & 2MASX  J07180060+4405271
 & Sy1 & 0.06 &  2.22$^{+0.41}_{-0.35}$ & 20.0&(2)\\ 
J0742.5+4947 & 115.644 & 49.793 & 2.98$\pm0.21$ & 14.4 & Mrk 79
 & Sy1.2 & 0.02 &  2.06$^{+0.17}_{-0.16}$ & 20.8&(1)\\ 
J0800.1+2322 & 120.032 & 23.370 & 1.62$\pm0.24$ & 6.6 & SDSS J0759.87+232448.3
 & GALAXY & 0.03 &  1.70$^{+0.31}_{-0.30}$ & 22.3&(2)\\ 
J0800.3+2638 & 120.099 & 26.648 & 1.79$\pm0.24$ & 7.5 & IC 486
 & Sy1 & 0.03 &  1.80$^{+0.30}_{-0.28}$ & 22.2&(2)\\ 
J0804.2+0506 & 121.050 & 5.101 & 3.18$\pm0.25$ & 12.6 & UGC 4203
 & Sy2 & 0.01 &  2.58$^{+0.60}_{-0.48}$ & 23.5&(10)\\ 
J0811.1+7602 & 122.798 & 76.049 & 1.15$\pm0.19$ & 6.2 & PG 0804+761
 & Sy1 & 0.10 &  2.58$^{+0.60}_{-0.48}$ & 20.0&(2)\\ 
J0814.4+0423 & 123.600 & 4.400 & 1.26$\pm0.24$ & 5.2 & CGCG 031-072
 & Sy1 & 0.03 &  1.90$^{+0.49}_{-0.42}$ & 23.3&(2)\\ 
J0823.2-0456 & 125.800 & -4.947 & 1.35$\pm0.23$ & 6.0 & SWIFT J0823.4-0457
 & Sy2 & 0.02 &  1.66$^{+0.38}_{-0.37}$ & 23.5&(3)\\ 
J0832.8+3706 & 128.200 & 37.100 & 1.03$\pm0.20$ & 5.2 & RBS 707
 & Sy1.2 & 0.09 &  2.41$^{+0.63}_{-0.52}$ & 20.0&(2)\\ 
J0839.8-1214 & 129.950 & -12.248 & 1.28$\pm0.21$ & 6.1 & 3C 206
 & Sy1 & 0.20 &  2.04$^{+0.34}_{-0.31}$ & 21.0&(2)\\ 
J0904.9+5537 & 136.250 & 55.632 & 1.04$\pm0.17$ & 6.0 & SWIFT J0904.3+5538
 & Sy1.5 & 0.04 &  1.92$^{+0.42}_{-0.39}$ & 21.0&(1)\\ 
J0911.5+4528 & 137.898 & 45.471 & 1.23$\pm0.18$ & 7.0 & SWIFT J0911.2+4533
 & Sy2 & 0.03 &  2.45$^{+0.54}_{-0.44}$ & 23.5&(1)\\ 
J0918.4+1618 & 139.615 & 16.316 & 1.65$\pm0.21$ & 8.0 & Mrk 704
 & Sy1.5 & 0.03 &  1.98$^{+0.26}_{-0.24}$ & 21.5 & (2)\\ 
J0921.0-0803 & 140.257 & -8.067 & 2.59$\pm0.20$ & 12.9 & SWIFT J0920.8-0805
 & Sy2 & 0.02 &  2.15$^{+0.21}_{-0.19}$ & 22.8&(1)\\ 
J0923.8+2256 & 140.962 & 22.936 & 2.05$\pm0.20$ & 10.5 & MCG +04-22-042
 & Sy1.2 & 0.03 &  1.85$^{+0.23}_{-0.21}$ & 20.6&(1)\\ 
J0925.2+5217 & 141.316 & 52.285 & 3.03$\pm0.17$ & 18.1 & Mrk 110
 & Sy1 & 0.04 &  2.00$^{+0.14}_{-0.13}$ & 20.6&(1)\\ 
J0945.8-1419 & 146.468 & -14.332 & 1.28$\pm0.21$ & 6.1 & NGC 2992
 & Sy2 & 0.01 &  1.55$^{+0.34}_{-0.33}$ & 22.0&(1)\\ 
J0947.7-3056 & 146.939 & -30.948 & 11.50$\pm0.22$ & 51.5 & ESO 434-40
 & Sy2 & 0.01 &  2.27$^{+0.01}_{-0.01}$ & 22.2&(3)\\ 
J0959.6-2250 & 149.916 & -22.834 & 4.44$\pm0.22$ & 19.8 & NGC 3081
 & Sy2 & 0.01 &  1.80$^{+0.13}_{-0.12}$ & 23.5&(1)\\ 
J1001.8+5542 & 150.453 & 55.700 & 1.43$\pm0.16$ & 8.8 & \bf{NGC 3079}
 & Sy2 & 0.004 &  1.88$^{+0.26}_{-0.25}$ & 24.7&(4)\\ 
J1006.0-2306 & 151.500 & -23.100 & 1.25$\pm0.23$ & 5.5 & ESO 499-G 041
 & Sy1 & 0.01 &  1.57$^{+0.60}_{-0.62}$ & 21.4&(2)\\ 
J1021.7-0327 & 155.450 & -3.450 & 1.35$\pm0.22$ & 6.3 & MCG+00-27-002
 & Sy1 & 0.04 &  2.34$^{+0.54}_{-0.44}$ & 20.0&(2)\\ 
J1023.5+1951 & 155.888 & 19.864 & 7.35$\pm0.20$ & 36.8 & NGC 3227
 & Sy1.5 & 0.004 &  1.98$^{+0.01}_{-0.01}$ & 22.8&(1)\\ 
J1031.8-3451 & 157.975 & -34.860 & 4.71$\pm0.26$ & 18.4 & NGC 3281
 & Sy2 & 0.01 &  1.98$^{+0.17}_{-0.16}$ & 23.9&(6)\\ 
J1031.9-1417 & 157.996 & -14.300 & 2.13$\pm0.23$ & 9.3 & H 1029-140
 & Sy1 & 0.09 &  2.17$^{+0.32}_{-0.29}$ & 20.0&(2)\\ 
J1044.0+7023 & 161.003 & 70.400 & 0.99$\pm0.17$ & 5.9 & MCG+12-10-067
 & Sy2 & 0.03 &  1.60$^{+0.37}_{-0.35}$ & 23.3&(2)\\ 
J1046.5+2556 & 161.649 & 25.950 & 1.12$\pm0.19$ & 5.9 & UGC 05881
 & GALAXY & 0.02 &  1.60$^{+0.38}_{-0.36}$ & 23.0&(2)\\ 
J1048.5-2512 & 162.149 & -25.200 & 1.42$\pm0.27$ & 5.3 & \bf{NGC 3393}
 & Sy2 & 0.01 &  2.01$^{+0.43}_{-0.37}$ & 24.7&(4)\\ 
J1049.3+2256 & 162.350 & 22.950 & 1.57$\pm0.20$ & 8.0 & SWIFT J1049.4+2258
 & Sy2 & 0.03 &  1.89$^{+0.24}_{-0.23}$ & 23.3&(2)\\ 
J1106.6+7234 & 166.654 & 72.571 & 6.45$\pm0.17$ & 38.0 & NGC 3516
 & Sy1.5 & 0.01 &  1.90$^{+0.01}_{-0.01}$ & 21.2&(1)\\ 
J1115.9+5426 & 168.999 & 54.450 & 0.88$\pm0.15$ & 5.7 & SDSS J111519.98+542316.6
 & Sy2 & 0.07 &  1.80$^{+0.57}_{-0.42}$ & 22.4&(2)\\ 
J1125.4+5421 & 171.352 & 54.351 & 0.97$\pm0.15$ & 6.3 & ARP 151
& Sy1 & 0.02 &  1.72$^{+0.48}_{-0.37}$ & 20.0&(2)\\ 
J1127.5+1908 & 171.900 & 19.148 & 1.12$\pm0.20$ & 5.6 & 1RXS J112716.6+190914
 & Sy1 & 0.10 &  1.84$^{+0.62}_{-0.54}$ & 21.4&(2)\\ 
J1132.7+5259 & 173.188 & 52.988 & 1.01$\pm0.15$ & 6.6 & UGC 6527
 & Sy1 & 0.03 &  1.85$^{+0.36}_{-0.34}$ & 20.0&(3)\\ 
J1136.5+2132 & 174.150 & 21.548 & 1.13$\pm0.19$ & 5.9 & Mrk 739
 & Sy1 & 0.03 &  3.06$^{+0.64}_{-0.53}$ & 20.7&(2)\\ 
J1139.0-3744 & 174.764 & -37.741 & 10.07$\pm0.27$ & 37.9 & NGC 3783
 & Sy1 & 0.01 &  1.94$^{+0.01}_{-0.01}$ & 22.5&(1)\\ 
J1139.1+5912 & 174.783 & 59.212 & 1.25$\pm0.15$ & 8.1 & SBS 1136+594
 & Sy1.5 & 0.06 &  2.76$^{+0.46}_{-0.39}$ & 19.6&(1)\\ 
J1139.4+3156 & 174.869 & 31.935 & 1.00$\pm0.17$ & 5.8 & NGC 3786
 & Sy1.8 & 0.01 &  1.74$^{+0.45}_{-0.34}$ & 22.5&(3)\\ 
J1144.7+7939 & 176.190 & 79.662 & 2.13$\pm0.18$ & 11.9 & SWIFT J1143.7+7942
 & Sy1.2 & 0.02 &  2.26$^{+0.24}_{-0.23}$ & 20.6&(1)\\ 
J1145.3+5859 & 176.349 & 59.000 & 0.81$\pm0.15$ & 5.3 & Ark 320
 & GALAXY & 0.01 &  2.23$^{+0.74}_{-0.55}$ & 22.2&(2)\\ 
J1145.5-1825 & 176.393 & -18.428 & 2.84$\pm0.27$ & 10.5 & 2MASX J11454045-1827149
 & Sy1 & 0.03 &  2.02$^{+0.22}_{-0.21}$ & 20.5&(1)\\ 
J1148.9+2938 & 177.230 & 29.634 & 1.06$\pm0.18$ & 6.1 & MCG+05-28-032
 & LINER & 0.02 &  1.85$^{+0.39}_{-0.35}$ & 22.5&(2)\\ 
J1158.0+5526 & 179.502 & 55.449 & 1.04$\pm0.15$ & 6.9 & NGC 3998
 & Sy1 & 0.003 &  2.05$^{+0.40}_{-0.41}$ & 20.1&(8)\\ 
J1201.0+0647 & 180.250 & 6.800 & 1.18$\pm0.21$ & 5.6 & SWIFT J1200.8+0650
 & Sy2 & 0.04 &  1.89$^{+0.33}_{-0.30}$ & 22.8&(1)\\ 
J1203.0+4432 & 180.773 & 44.534 & 2.33$\pm0.15$ & 15.1 & NGC 4051
 & Sy1.5 & 0.002 &  2.45$^{+0.18}_{-0.17}$ & 20.5&(1)\\ 
J1204.5+2018 & 181.149 & 20.301 & 1.32$\pm0.19$ & 7.1 & ARK 347
 & Sy2 & 0.02 &  1.76$^{+0.29}_{-0.23}$ & 23.2&(1)\\ 
J1206.2+5242 & 181.565 & 52.710 & 1.24$\pm0.15$ & 8.3 & NGC 4102
 & LINER & 0.003 &  1.74$^{+0.30}_{-0.28}$ & 20.9&(1)\\ 
J1209.1+4700 & 182.300 & 47.000 & 0.76$\pm0.15$ & 5.0 & Mrk 198
 & Sy2 & 0.02 &  1.76$^{+0.47}_{-0.36}$ & 22.8&(2)\\ 
J1209.4+4341 & 182.370 & 43.686 & 1.56$\pm0.15$ & 10.1 & NGC 4138
 & Sy1.9 & 0.003 &  1.90$^{+0.22}_{-0.21}$ & 22.9&(1)\\ 
J1210.5+3924 & 182.633 & 39.406 & 24.60$\pm0.16$ & 153.4 & NGC 4151
 & Sy1.5 & 0.003 &  1.93$^{+0.01}_{-0.01}$ & 22.5&(1)\\ 
J1210.6+3819 & 182.667 & 38.333 & 0.96$\pm0.16$ & 6.0 & LEDA 38759
 & Sy1 & 0.02 &  1.91$^{+0.35}_{-0.33}$ & 22.6&(2)\\ 
J1217.2+0711 & 184.300 & 7.200 & 1.21$\pm0.20$ & 5.9 & NGC 4235
 & Sy1 & 0.01 &  1.59$^{+0.44}_{-0.44}$ & 21.2&(3)\\ 
J1218.3+2950 & 184.593 & 29.839 & 1.53$\pm0.17$ & 9.1 & Mrk 766
 & Sy1.5 & 0.01 &  3.07$^{+0.42}_{-0.36}$ & 21.7&(1)\\ 
J1219.0+4715 & 184.750 & 47.252 & 0.96$\pm0.15$ & 6.3 & NGC 4258
 & Sy1 & 0.001 &  1.92$^{+0.37}_{-0.34}$ & 22.9&(11)\\ 
J1222.0+7518 & 185.503 & 75.311 & 1.27$\pm0.17$ & 7.4 & Mrk 205
 & Sy1 & 0.07 &  2.53$^{+0.42}_{-0.37}$ & 20.7&(12)\\ 
J1225.7+1239 & 186.447 & 12.665 & 12.58$\pm0.19$ & 65.6 & NGC 4388
 & Sy2 & 0.01 &  1.84$^{+0.01}_{-0.01}$ & 23.6&(1)\\ 
J1225.8+3330 & 186.466 & 33.513 & 1.25$\pm0.16$ & 7.7 & NGC 4395
 & Sy1 & 0.001 &  2.15$^{+0.27}_{-0.25}$ & 22.3&(1)\\ 
J1235.6-3955 & 188.902 & -39.919 & 10.21$\pm0.26$ & 39.4 & NGC 4507
 & Sy2 & 0.01 &  1.98$^{+0.19}_{-0.14}$ & 23.5&(1)\\ 
J1238.8-2718 & 189.723 & -27.308 & 4.39$\pm0.28$ & 15.9 & ESO 506-027
 & Sy2 & 0.02 &  1.74$^{+0.13}_{-0.13}$ & 23.6&(9)\\ 
J1239.0-1611 & 189.769 & -16.196 & 2.02$\pm0.26$ & 7.7 & XSS J12389-1614
 & Sy2 & 0.04 &  1.68$^{+0.28}_{-0.22}$ & 22.5&(1)\\ 
J1239.5-0520 & 189.898 & -5.341 & 4.52$\pm0.23$ & 19.8 & NGC 4593
 & Sy1 & 0.01 &  1.91$^{+0.11}_{-0.11}$ & 20.3&(1)\\ 
J1246.6+5434 & 191.661 & 54.575 & 1.34$\pm0.15$ & 9.0 & NGC 4686
 & LINER & 0.02 &  1.75$^{+0.25}_{-0.20}$ & 23.8&(2)\\ 
J1302.8+1624 & 195.700 & 16.400 & 0.90$\pm0.17$ & 5.1 & Mrk 0783
 & Sy1.2 & 0.07 &  1.92$^{+0.43}_{-0.41}$ & 21.0&(2)\\ 
J1306.7-4024 & 196.698 & -40.415 & 2.37$\pm0.27$ & 8.9 & ESO 323-077
 & Sy1.2 & 0.02 &  2.03$^{+0.27}_{-0.25}$ & 22.7&(2)\\ 
J1309.1+1137 & 197.279 & 11.632 & 2.19$\pm0.18$ & 12.0 & SWIFT J1309.2+1139
 & Sy2 & 0.03 &  1.63$^{+0.15}_{-0.15}$ & 23.4&(13)\\ 
J1315.4+4424 & 198.852 & 44.404 & 1.28$\pm0.15$ & 8.4 & IGR J13149+4422
 & Sy & 0.04 &  2.28$^{+0.31}_{-0.28}$ & 22.8&(2)\\ 
J1322.3-1642 & 200.591 & -16.716 & 2.57$\pm0.27$ & 9.5 & MCG -03-34-064
 & Sy1.8 & 0.02 &  2.15$^{+0.30}_{-0.28}$ & 23.6&(1)\\ 
J1325.4-4301 & 201.366 & -43.017 & 49.77$\pm0.26$ & 187.6 & Cen A
 & Sy2 & 0.002 &  1.85$^{+0.00}_{-0.00}$ & 22.7&(1)\\ 
J1334.8-2323 & 203.700 & -23.400 & 1.49$\pm0.29$ & 5.1 & ESO 509-38
 & Sy2 & 0.03 &  2.37$^{+0.69}_{-0.54}$ & 20.0&(2)\\ 
J1335.7-3418 & 203.944 & -34.302 & 4.86$\pm0.29$ & 16.6 & MCG -06-30-015
 & Sy1.2 & 0.01 &  2.24$^{+0.17}_{-0.17}$ & 21.7&(1)\\ 
J1338.1+0433 & 204.547 & 4.552 & 3.63$\pm0.20$ & 17.9 & NGC 5252
 & Sy2 & 0.02 &  1.67$^{+0.12}_{-0.12}$ & 22.64&(13)\\ 
J1341.4+3022 & 205.356 & 30.369 & 1.15$\pm0.16$ & 7.2 & Mrk 268
 & Sy2 & 0.04 &  2.38$^{+0.35}_{-0.30}$ & 23.3&(2)\\ 
J1349.5-3018 & 207.390 & -30.304 & 17.87$\pm0.31$ & 58.2 & IC 4329A
 & Sy1 & 0.02 &  2.05$^{+0.00}_{-0.00}$ & 21.6&(1)\\ 
J1353.2+6919 & 208.305 & 69.327 & 2.78$\pm0.17$ & 16.7 & Mrk 279
 & Sy1.5 & 0.03 &  1.96$^{+0.15}_{-0.15}$ & 20.5&(1)\\ 
J1356.1+3835 & 209.033 & 38.583 & 1.22$\pm0.16$ & 7.7 & Mrk 464
 & Sy1 & 0.05 &  1.69$^{+0.30}_{-0.29}$ & 20.0&(3)\\ 
J1408.4-3024 & 212.100 & -30.400 & 1.65$\pm0.32$ & 5.1 & PGC 050427
 & Sy1 & 0.02 &  2.67$^{+0.65}_{-0.52}$ & 21.2&(2)\\ 
J1413.5-0312 & 213.375 & -3.201 & 14.38$\pm0.24$ & 59.0 & NGC 5506
 & Sy1.9 & 0.01 &  2.27$^{+0.11}_{-0.14}$ & 22.5&(1)\\ 
J1418.2+2507 & 214.568 & 25.133 & 3.12$\pm0.17$ & 18.2 & NGC 5548
 & Sy1.5 & 0.02 &  1.82$^{+0.12}_{-0.12}$ & 20.4&(1)\\ 
J1419.5-2639 & 214.893 & -26.663 & 3.49$\pm0.34$ & 10.4 & ESO 511-G030
 & Sy1 & 0.02 &  2.11$^{+0.22}_{-0.20}$ & 21.2&(1)\\ 
J1421.6+4750 & 215.420 & 47.838 & 1.03$\pm0.16$ & 6.4 & QSO B1419+480
 & Sy1 & 0.07 &  1.73$^{+0.35}_{-0.27}$ & 21.3&(3)\\ 
J1424.3+2435 & 216.100 & 24.600 & 0.89$\pm0.17$ & 5.1 & NGC 5610
 & GALAXY & 0.02 &  1.88$^{+0.46}_{-0.42}$ & 22.8&(2)\\ 
J1429.6+0117 & 217.400 & 1.300 & 1.26$\pm0.23$ & 5.4 & QSO B1426+015
 & Sy1 & 0.09 &  2.41$^{+0.74}_{-0.59}$ & 20.0&(3)\\ 
J1436.5+5847 & 219.149 & 58.798 & 1.37$\pm0.16$ & 8.3 & QSO J1436+5847
 & Sy1 & 0.03 &  1.68$^{+0.27}_{-0.22}$ & 23.5&(2)\\ 
J1441.2+5330 & 220.300 & 53.500 & 0.85$\pm0.16$ & 5.1 & Mrk 477
 & Sy2 & 0.04 &  1.47$^{+0.88}_{-0.46}$ & 24.0&(14)\\ 
J1442.6-1713 & 220.664 & -17.223 & 4.83$\pm0.34$ & 14.4 & \bf{NGC 5728}
 & Sy2 & 0.01 &  1.84$^{+0.14}_{-0.14}$ & 24.3&(4)\\ 
J1453.1+2556 & 223.282 & 25.936 & 1.29$\pm0.18$ & 7.0 & RX J1453.1+2554
 & Sy1 & 0.05 &  1.82$^{+0.26}_{-0.25}$ & 20.0&(2)\\ 
J1504.2+1025 & 226.073 & 10.417 & 1.51$\pm0.22$ & 6.8 & Mrk 841
 & Sy1 & 0.04 &  1.80$^{+0.38}_{-0.36}$ & 21.3&(1)\\ 
J1515.4+4201 & 228.868 & 42.033 & 1.05$\pm0.18$ & 5.9 & NGC 5899
 & Sy2 & 0.01 &  1.72$^{+0.56}_{-0.51}$ & 23.1&(2)\\ 
J1536.2+5753 & 234.061 & 57.890 & 1.43$\pm0.18$ & 7.9 & Mrk 290
 & Sy1 & 0.03 &  2.19$^{+0.39}_{-0.34}$ & 20.4&(1)\\ 
J1548.4-1344 & 237.106 & -13.749 & 2.91$\pm0.39$ & 7.4 & NGC 5995
 & Sy2 & 0.03 &  2.01$^{+0.26}_{-0.24}$ & 22.0&(2)\\ 
J1554.8+3242 & 238.700 & 32.700 & 1.04$\pm0.20$ & 5.2 & 2MASX J15541741+3238381
 & Sy1 & 0.05 &  1.66$^{+0.76}_{-0.52}$ & 20.0&(2)\\ 
J1628.3+5147 & 247.082 & 51.793 & 2.45$\pm0.20$ & 12.4 & SWIFT J1628.1+5145
 & Sy1.9 & 0.05 &  2.13$^{+0.19}_{-0.18}$ & 23.3&(1)\\ 
J1653.2+0224 & 253.319 & 2.404 & 4.20$\pm0.35$ & 12.1 & \bf{NGC 6240}
 & Sy2 & 0.02 &  2.09$^{+0.18}_{-0.17}$ & 24.3&(4)\\ 
J1822.1+6421 & 275.541 & 64.361 & 1.10$\pm0.21$ & 5.3 & QSO B1821+643
 & Sy1 & 0.30 &  2.91$^{+0.97}_{-0.65}$ & 20.0&(15)\\ 
J1824.2-5620 & 276.057 & -56.348 & 1.98$\pm0.29$ & 6.9 & IC 4709
 & Sy2 & 0.02 &  1.94$^{+0.28}_{-0.26}$ & 23.1&(2)\\ 
J1835.1+3240 & 278.791 & 32.683 & 4.67$\pm0.21$ & 22.0 & 3C 382
 & Sy1 & 0.06 &  2.09$^{+0.11}_{-0.10}$ & 21.1&(1)\\ 
J1837.1-5922 & 279.284 & -59.368 & 1.79$\pm0.28$ & 6.3 & FAIRALL 49
 & Sy2 & 0.02 &  2.55$^{+0.43}_{-0.37}$ & 22.3&(14)\\ 
J1838.6-6523 & 279.658 & -65.394 & 6.23$\pm0.28$ & 22.4 & ESO 103-035
 & Sy2 & 0.01 &  2.18$^{+0.12}_{-0.12}$ & 23.2&(1)\\ 
J1842.4+7946 & 280.616 & 79.771 & 5.82$\pm0.19$ & 29.9 & 3C 390.3
 & Sy1 & 0.06 &  1.95$^{+0.07}_{-0.07}$ & 21.0&(1)\\ 
J1845.1-6223 & 281.297 & -62.399 & 2.52$\pm0.28$ & 9.0 & ESO 140-43
 & Sy1 & 0.01 &  2.17$^{+0.27}_{-0.25}$ & 22.4&(2)\\ 
J1857.3-7827 & 284.341 & -78.464 & 1.86$\pm0.26$ & 7.1 & LEDA 140831
 & Sy1 & 0.04 &  1.85$^{+0.36}_{-0.34}$ & 20.0&(2)\\ 
J1921.2-5840 & 290.323 & -58.677 & 3.54$\pm0.28$ & 12.6 & ESO 141-55
 & Sy1 & 0.04 &  1.83$^{+0.19}_{-0.18}$ & 20.0&(3)\\ 
J1942.7-1018 & 295.680 & -10.316 & 4.30$\pm0.30$ & 14.3 & NGC 6814
 & Sy1 & 0.01 &  1.91$^{+0.15}_{-0.14}$ & 20.8&(1)\\ 
J2009.1-6103 & 302.289 & -61.064 & 3.03$\pm0.26$ & 11.5 & SWIFT J2009.0-6103
 & Sy1 & 0.01 &  1.98$^{+0.21}_{-0.20}$ & 21.8&(1)\\ 
J2044.1-1043 & 311.039 & -10.731 & 5.62$\pm0.29$ & 19.7 & Mrk 509
 & Sy1.2 & 0.03 &  2.04$^{+0.12}_{-0.11}$ & 20.7&(1)\\ 
J2052.0-5703 & 313.017 & -57.063 & 4.63$\pm0.25$ & 18.3 & IC 5063
 & Sy2 & 0.01 &  1.89$^{+0.12}_{-0.12}$ & 23.3&(1)\\ 
J2109.1-0939 & 317.300 & -9.652 & 1.58$\pm0.27$ & 5.9 & 1H 2107-097
 & Sy1 & 0.03 &  2.15$^{+0.40}_{-0.35}$ & \nodata\\ 
J2132.1-3343 & 323.028 & -33.727 & 2.92$\pm0.27$ & 10.7 & CTS 109
 & Sy1 & 0.03 &  2.07$^{+0.22}_{-0.20}$ & 20.0&(2)\\ 
J2136.0-6223 & 324.006 & -62.400 & 2.42$\pm0.23$ & 10.6 & QSO J2136-6224
 & Sy1 & 0.06 &  2.23$^{+0.26}_{-0.24}$ & 20.0&(2)\\ 
J2138.8+3206 & 324.713 & 32.115 & 1.29$\pm0.00$ & 6.3 & LEDA 67084
 & Sy1 & 0.03 &  2.06$^{+0.55}_{-0.48}$ & 20.0&(2)\\ 
J2200.7+1033 & 330.199 & 10.565 & 1.76$\pm0.21$ & 8.6 & SWIFT J2200.9+1032
 & Sy1.9 & 0.03 &  2.08$^{+0.27}_{-0.25}$ & 22.2&(1)\\ 
J2202.1-3152 & 330.526 & -31.878 & 8.01$\pm0.25$ & 31.8 & NGC 7172
 & Sy2 & 0.01 &  1.80$^{+0.01}_{-0.01}$ & 22.9&(1)\\ 
J2204.5+0335 & 331.149 & 3.600 & 1.33$\pm0.21$ & 6.3 & IRAS 22017+0319
 & Sy2 & 0.06 &  2.29$^{+0.48}_{-0.40}$ & 22.5&(2)\\ 
J2209.5-4709 & 332.387 & -47.166 & 3.02$\pm0.23$ & 13.4 & NGC 7213
 & Sy1.5 & 0.03 &  1.92$^{+0.17}_{-0.16}$ & 20.6&(1)\\ 
J2223.8-0207 & 335.962 & -2.121 & 1.93$\pm0.22$ & 8.9 & 3C 445
 & Sy1 & 0.06 &  1.99$^{+0.24}_{-0.23}$ & 23.2&(2)\\ 
J2235.8-2603 & 338.966 & -26.054 & 2.76$\pm0.24$ & 11.6 & NGC 7314
 & Sy1.9 & 0.005 &  2.04$^{+0.20}_{-0.18}$ & 21.8&(1)\\ 
J2236.1+3357 & 339.040 & 33.952 & 1.66$\pm0.19$ & 8.8 & Arp 319
 & Sy2 & 0.02 &  1.75$^{+0.27}_{-0.23}$ & 23.7&(2)\\ 
J2236.8-1235 & 339.223 & -12.599 & 1.43$\pm0.23$ & 6.2 & Mrk 915
 & Sy1 & 0.02 &  1.59$^{+0.33}_{-0.27}$ & 22.8&(2)\\ 
J2245.7+3941 & 341.449 & 39.695 & 1.71$\pm0.18$ & 9.2 & 3C 452
 & Sy2 & 0.08 &  1.61$^{+0.24}_{-0.19}$ & 23.4&(1)\\ 
J2254.1-1734 & 343.535 & -17.578 & 5.67$\pm0.23$ & 24.7 & MR 2251-178
 & Sy1 & 0.06 &  2.06$^{+0.13}_{-0.12}$ & 20.8&(1)\\ 
J2258.9+4053 & 344.749 & 40.899 & 1.31$\pm0.18$ & 7.2 & UGC 12282
 & Sy1 & 0.02 &  1.65$^{+0.28}_{-0.22}$ & 23.9&(2)\\ 
J2259.5+2455 & 344.899 & 24.929 & 1.51$\pm0.19$ & 8.0 & LEDA 70195
 & Sy1 & 0.03 &  1.86$^{+0.29}_{-0.28}$ & 20.0&(2)\\ 
J2303.2+0853 & 345.809 & 8.885 & 3.87$\pm0.20$ & 19.4 & NGC 7469
 & Sy1.2 & 0.02 &  2.14$^{+0.12}_{-0.12}$ & 20.6&(1)\\ 
J2304.7-0841 & 346.194 & -8.686 & 6.09$\pm0.22$ & 27.9 & Mrk 926
 & Sy1.5 & 0.05 &  1.97$^{+0.01}_{-0.01}$ & 21.1&(1)\\ 
J2304.7+1217 & 346.200 & 12.300 & 1.11$\pm0.20$ & 5.6 & NGC 7479
 & Sy2 & 0.01 &  1.98$^{+0.40}_{-0.37}$ & 23.6&(3)\\ 
J2318.4-4221 & 349.614 & -42.360 & 4.09$\pm0.20$ & 20.7 & \bf{NGC 7582}
 & Sy2 & 0.01 &  1.95$^{+0.22}_{-0.24}$ & 24.1&(4)\\ 
J2319.0+0014 & 349.762 & 0.241 & 2.82$\pm0.00$ & 13.5 & NGC 7603
 & Sy1 & 0.03 &  2.07$^{+0.18}_{-0.17}$ & 20.0&(2)\\ 
J2326.3+2154 & 351.600 & 21.900 & 0.97$\pm0.19$ & 5.1 & RBS 2005
 & Sy1 & 0.12 &  2.05$^{+0.52}_{-0.47}$ & 20.3&(2)\\ 
J2342.0+3035 & 355.500 & 30.600 & 1.08$\pm0.19$ & 5.7 & UGC 12741
 & Sy2 & 0.02 &  1.64$^{+0.33}_{-0.30}$ & 23.7&(2)\\ 

\enddata


\tablerefs{ (1) \citealp{tueller08}; 
(2) This work: follow up with XRT; 
(3) This work: follow up with XMM; 
(4) See Tab.~\ref{tab:cthick} for a detailed analysis of the Compton-thick sources. 
(5) \citealp{shinozaki06};
(6) \citealp{winter09};
(7) \citealp{ueda07};
(8) \citealp{georga05};
(9) \citealp{winter09b};
(10) \citealp{matt09}; 
(11) \citealp{cappi06};
(12) \citealp{page05};
(13) \citealp{winter08};
(14) \citealp{shu07};
(15) \citealp{jimenez07}.
}
\item[$^{\mathrm{\ddag}}$] The optical classification comes mainly from Tueller et al. (2008), Winter et al. (2009a), \cite{parisi09}, \cite{cusumano10}, SIMBAD, and NED.

\end{deluxetable}

\end{document}